\documentclass[review, times, 12pt]{elsarticle}
\def\imgdir{}
\usepackage[margin=1in]{geometry}
\usepackage{setspace}
\usepackage{lineno} 
\usepackage{amssymb}
\usepackage{amsfonts}
\usepackage{amsmath}
\usepackage{amsthm}
\usepackage{txfonts}
\usepackage{textcomp}
\usepackage{graphicx}
\usepackage[english]{babel}
\usepackage{epsfig}
\usepackage{subcaption}
\usepackage{xcolor}
\usepackage{color,soul}
\usepackage{todonotes}
\usepackage{booktabs}
\usepackage{pifont}
\usepackage{multirow}
\usepackage{diagbox}
\usepackage{mathtools}
\usepackage{lipsum}
\usepackage{wrapfig}
\usepackage{hyperref} 

\journal{Additive Manufacturing}

\begin{document}
\begin{frontmatter}
\title{Airborne acoustic emission enables sub-scanline keyhole porosity quantification and effective process characterization for metallic laser powder bed fusion}

\author[label1,label2,label3]{Haolin~Liu\corref{cor1}}
\author[label1,label3]{David~Guirguis\corref{cor2}}
\author[label1]{Xuzhe~Zeng\corref{cor2}}
\author[label1,label3]{Logan~Maurer}
\author[label1]{Vigknesh~Rajan}
\author[label4]{Niloofar~Sanaei}
\author[label4]{Chi-Ta~Yang}z
\author[label1,label2,label3]{Jack~L.~Beuth\corref{cor1}}
\author[label2,label3]{Anthony~D.~Rollett\corref{cor1}}
\author[label1,label3]{Levent~Burak~Kara\corref{cor1}}

\address[label1]{Department of Mechanical Engineering, Carnegie Mellon University, Pittsburgh, Pennsylvania, USA}

\address[label2]{Department of Materials Science and Engineering, Carnegie Mellon University, Pittsburgh, Pennsylvania, USA}

\address[label3]{NextManufacturing Center, Carnegie Mellon University, Pittsburgh, Pennsylvania, USA}

\address[label4]{Digital Engineering and Advanced Manufacturing, Eaton Corporation, Southfield, MI 48076, USA}

\cortext[cor1]{Corresponding authors. \\\hspace*{1.6em} E-mail address: \{haolinl, beuth, rollett, lkara\}@andrew.cmu.edu}

\cortext[cor2]{These authors contributed equally to this work.}


\begin{abstract}
{
Keyhole-induced (KH) porosity, which arises from unstable vapor cavity dynamics under excessive laser energy input, remains a significant challenge in laser powder bed fusion (LPBF). This study presents an integrated experimental and data-driven framework using airborne acoustic emission (AE) to achieve high-resolution quantification of KH porosity. Experiments conducted on an LPBF system involved \textit{in~situ} acquisition of airborne AE and \textit{ex~situ} porosity imaging via X-ray computed tomography (XCT), synchronized spatiotemporally through photodiode signals with submillisecond precision. We introduce $\texttt{KHLineNum}$, a spatially resolved porosity metric defined as the number of KH pores per unit scan length, which serves as a physically meaningful indicator of the severity of KH porosity in geometries and scanning strategies. Using AE scalogram data and scan speed, we trained a lightweight convolutional neural network to predict $\texttt{KHLineNum}$ with millisecond-scale temporal resolution, achieving an $R^2$ value exceeding 0.8. Subsequent analysis identified the $35{-}45~\mathrm{kHz}$ frequency band of AE as particularly informative, consistent with known KH oscillations. Beyond defect quantification, the framework also enables AE-driven direct inference of KH regime boundaries on the power–velocity process map, offering a noninvasive and scalable component to labor-intensive post-process techniques such as XCT. We believe this framework advances AE-based monitoring in LPBF, providing a pathway toward improved quantifiable defect detection and process control. 
}
\end{abstract}

\begin{graphicalabstract}
\label{sec:graphical}
\begin{figure*}[!ht]
\center{\includegraphics[width=\linewidth]
{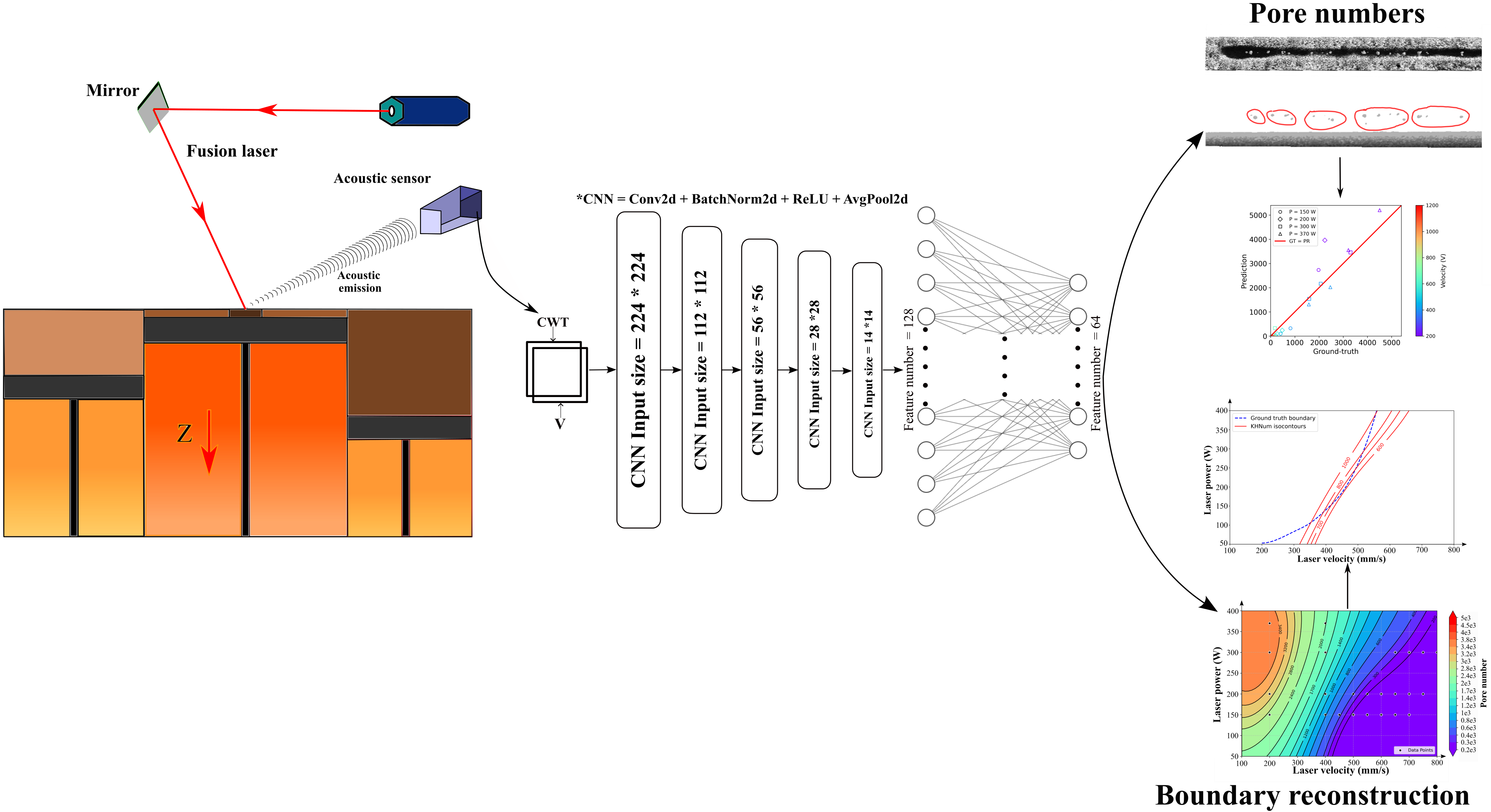}
}
\end{figure*}
\end{graphicalabstract}

\begin{highlights}
\item Novel defect metric---$\texttt{KHLineNum}$: Introduces $\texttt{KHLineNum}$, a linear number density of keyhole-induced (KH) pores per unit scan length, offering a physically grounded spatially resolved measure of KH porosity severity, which is shown to be more informative than volumetric or total pore count.
\item Sub-Scanline KH porosity quantification via acoustic emission (AE): Developed a framework that synchronizes airborne AE signals with KH porosity data via the \textit{ex~situ} X-ray computed tomography (XCT) scanning at millisecond resolution to localize KH pore formation events along the scanline. Trained a 2-channel machine learning model (AE scalogram + scan speed ($V$) to predict $\texttt{KHLineNum}$ with $R^2>0.8$, enabling fine-grained, quantitative AE-based inference of KH porosity. The predictions of the non-overlapping sliding windows demonstrated the potential of the framework for future integration into online monitoring systems.
\item AE frequency band sensitivity: Identified $37\mathrm{-}42~\mathrm{kHz}$ as the most informative AE frequency range for KH pore prediction, which is consistent with prior studies on keyhole oscillation dynamics. 
\item Cost-effective, non-destructive framework based on AE for process optimization: Enabled inference of KH process window boundaries (KH-bounds) directly from AE and $V$ and effectively reproduced a known KH-bound on the power–velocity map. Offers a scalable alternative to XCT for KH porosity quantification and process map characterization, with potential for \textit{in~situ} process control in industrial LPBF systems.
\end{highlights}

\begin{keyword}
{
Laser powder bed fusion (LPBF); keyhole-induced (KH) defect quantification; airborne acoustic emission (AE) monitoring; linear number density of KH porosity ($\texttt{KHLineNum}$); KH regime boundary (KH-bound) characterization 
}
\end{keyword}

\end{frontmatter}


\section{Introduction}
\label{sec:intro}
Laser powder bed fusion (LPBF) has emerged as a cornerstone of metal additive manufacturing (AM), enabling the production of geometrically complex, high-performance components with excellent material efficiency and design flexibility~\cite{gordon2020defect}. However, the quality of LPBF parts is often compromised by process-associated porosity, which degrades density, strength, ductility, and fatigue life~\cite{kan2022critical,guo2023identifying,mower2016mechanical,edwards2014fatigue}. Porosity in LPBF arises predominantly from three mechanisms: lack of fusion (LoF) due to insufficient melting and spatters, bead-up (BU) caused by surface-tension-driven balling, and keyhole-induced (KH) porosity resulting from excessive laser energy input. KH porosity, the central focus of this study, occurs when intense laser irradiation vaporizes the metal, forming a high-pressure cavity that can collapse and entrap spherical voids~\cite{guo2023identifying,ren2024sub}. As KH porosity is a critical and prevalent defect mode in high-energy LPBF regimes, its detection and mitigation are essential for ensuring the structural integrity and reliability of AM parts.

To characterize porosity and related defects in LPBF, a variety of characterization techniques have been developed, both \textit{in~situ} and \textit{ex~situ}, including high-speed melt pool/plume imaging, build plate monitoring, acoustic emission (AE) sensing, and X-ray computed tomography (XCT)~\cite{du2019effects,forien2020detecting,wahlquist2024roles,chen2023situ,tempelman2022detection,tempelman2022sensor}. Among these, XCT is widely recognized for its ability to generate high-resolution, three-dimensional (3-D) images of internal defects, making it a valuable tool for post-process part quality evaluation. However, its application is limited by high operational costs, long acquisition times, and the lack of prompt feedback, since the scans are performed mostly with high resolution and after the completion of the build~\cite{guillen2024critical}. This post~hoc nature eliminates XCT's utility for adaptive control during fabrication. In contrast, AE sensing has emerged as a promising \textit{in~situ} monitoring approach, capable of capturing airborne acoustic waves spanning a broad frequency range that are generated by the complex laser–material interactions during LPBF~\cite{tempelman2024uncovering, drissi2022differentiation, shevchik2018acoustic, shevchik2019deep}. By continuously detecting and analyzing the emission signals, prior studies have demonstrated AE's feasibility of inferring adverse fusion behaviors and capturing defect formation in real time~\cite{liu2024inference,hamidi2023harmonizing}, highlighting AE’s potential for defect detection and multimodal process monitoring of more complex phenomena. 

The integration of AE data with machine learning (ML) has significantly advanced \textit{in~situ} process monitoring in LPBF, offering a promising path toward intelligent, data-driven control built on inherently noisy process signals. Deep learning models, including convolutional neural networks (CNN) and deep residual networks (ResNet), have been employed to extract patterns from AE signals and predict key process information~\cite{wasmer2018situ,pandiyan2022deep}. However, most studies frame AE-based porosity detection as a classification task rather than direct quantitative estimation. For example, CNNs trained on AE-derived spectrograms have been used to classify LPBF processes into categories of KH porosity, conduction mode, and LoF porosity~\cite{drissi2023towards}, while Liu et al.~\cite{liu2021physics} developed a model to categorize parts into semantic quality labels (\textit{i.e.}, pass, flag, fail) based on porosity levels. Although classification provides qualitative insight into defect detection, it lacks the resolution for accurate porosity quantification required for accurate online process characterization and optimization. Liu et al.~\cite{liu2024inference} proposed a ResNet-based framework to predict the geometry of the melt pool and infer the LoF porosity from AE signals but there is still a lack of a regression model that directly quantifies the KH porosity. One of the central challenges in achieving this goal is identifying a KH porosity measure that not only captures the severity of porosity formation but also aligns closely with the characteristics extracted from airborne AE signals. Discovering such a measure would facilitate quantification of KH porosity and, in turn, enable finer resolution in defect monitoring and more accurate control over the LPBF process.  

\begin{figure}[!t]
\center{\includegraphics[width=\linewidth]
{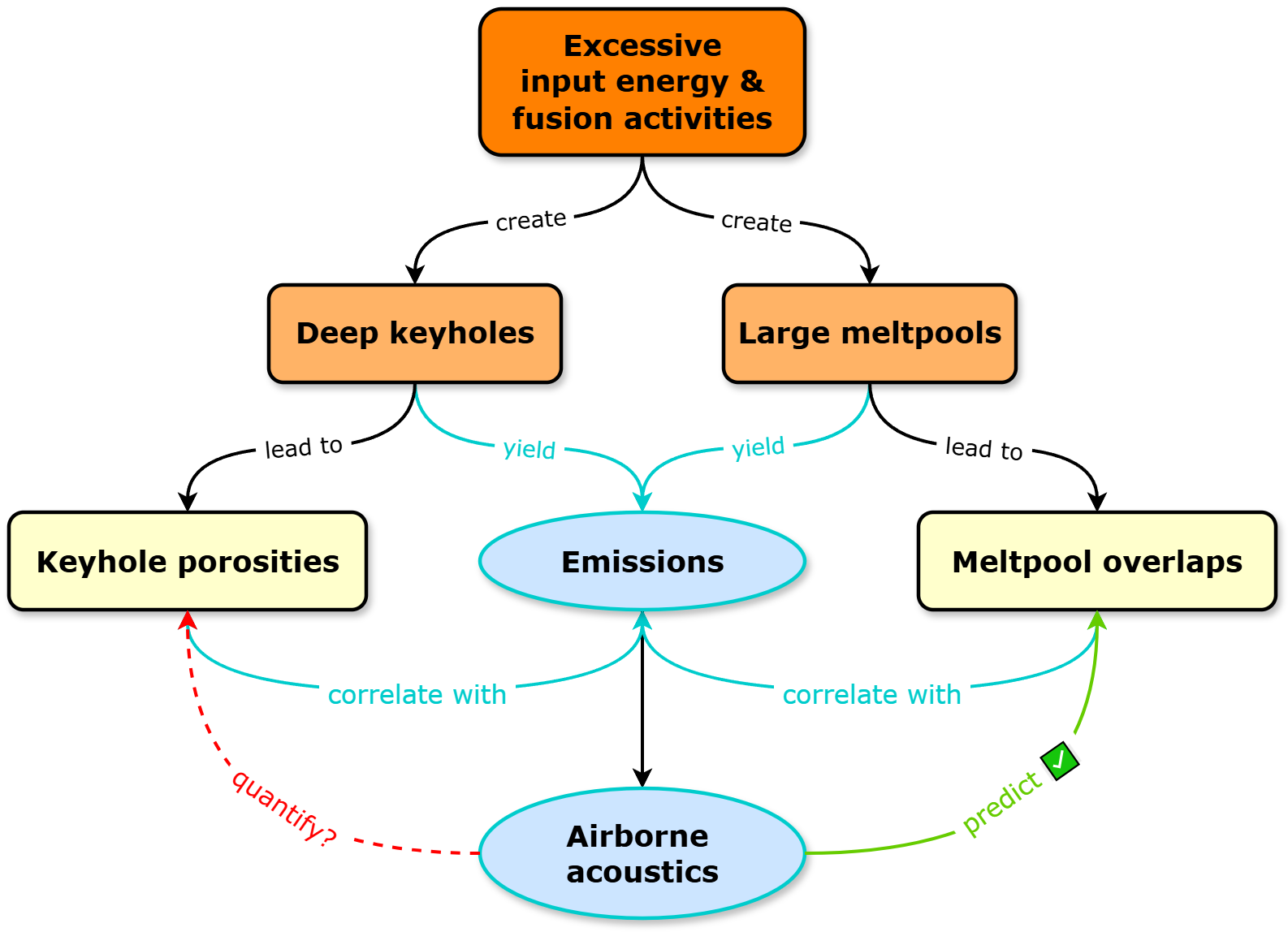}
\caption{Rationale and motivation for acoustic-based KH porosity quantification. The green line represents one of our previous works~\cite{liu2024inference}, and the red dashed line indicates the emphasis of this study. }
\label{fig:flowchart}
}
\end{figure}

As shown in Fig.~\ref{fig:flowchart}, this study addresses the gap in quantitative porosity estimation by investigating the ability of airborne AE signals, combined with process parameters, to predict the spatial density of KH porosities in LPBF-fabricated metal parts. The approach is based on controlled bare-plate experiments conducted on an EOS M290 system, where \textit{in~situ} AE data and \textit{ex~situ} XCT-based porosity measurements were collected and accurately aligned in time and space. After calibration and synchronization, the AE data were transformed into corresponding spectrotemporal representations, called scalogram snippets, via spectral analysis. These AE scalograms snippets were subsequently paired with process parameters and a derived KH porosity metric, $\texttt{KHLineNum}$, which represents the number of KH pores generated per unit length of scan. A CNN trained on this dataset achieved millisecond-level porosity quantification and generalized well across scan geometries and patterns. Subsequent ablation studies identified the most informative frequency bands for prediction. The framework also enables cost-efficient inference of KH regime boundaries on the power–velocity ($P\mathrm{-}V$) process map, offering a path toward AE-driven process characterization. By upgrading from classification to regression, this work demonstrates that AE signals can support high-resolution quantification of KH porosity, advancing AE-based LPBF monitoring and laying the groundwork for future control strategies.

\section{Methodology}
\label{sec:method}
\begin{figure*}[!ht]
\center{\includegraphics[width=\linewidth]
{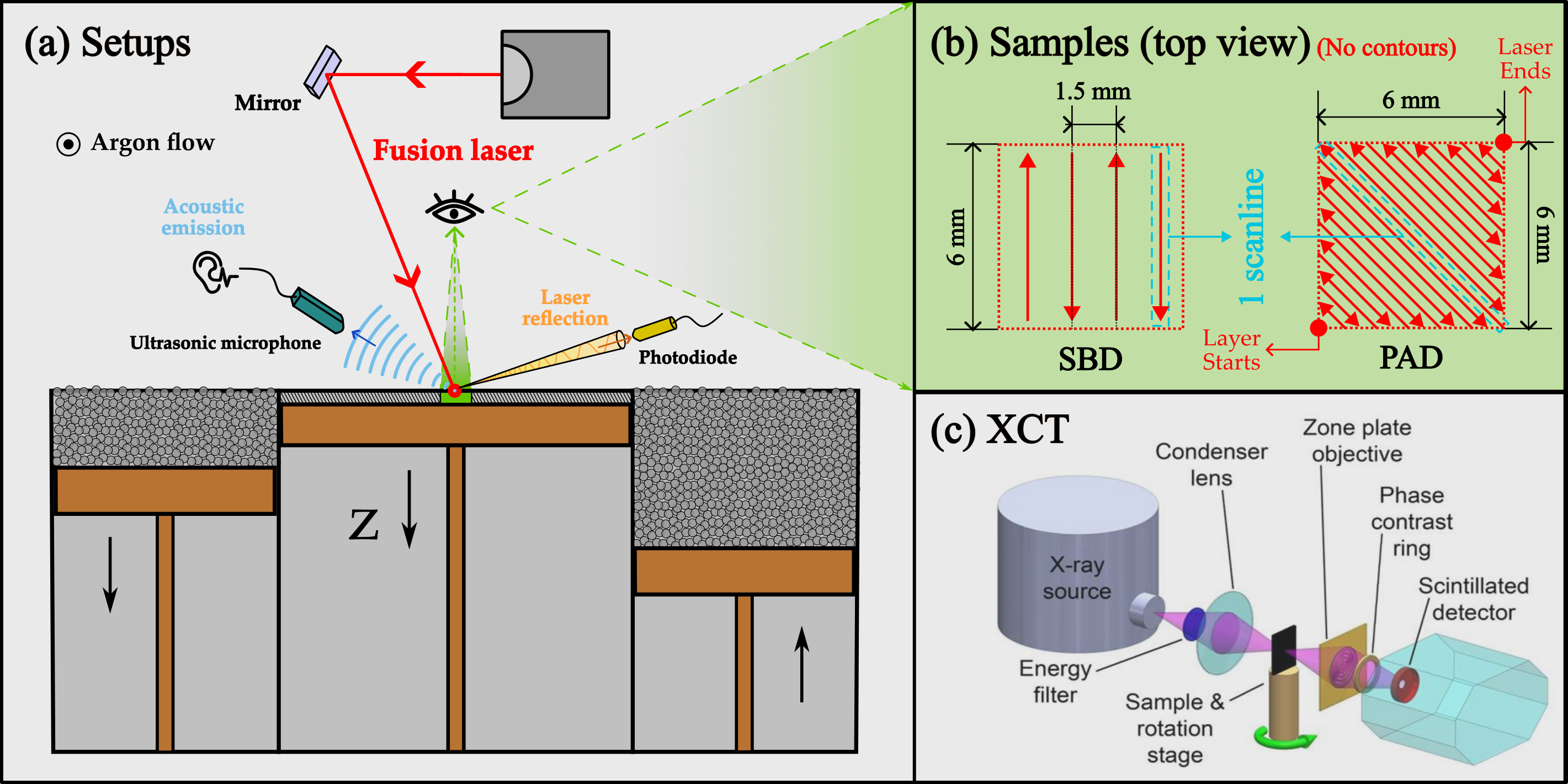}
\caption{An overview of experimental details. (a) LPBF machine and the associated sensor setups. (b) Details of the SBD and PAD samples, both of which consist of multiple scanlines and contain no contours. (c) XCT equipment used for \textit{ex~situ} KH porosity extraction~\cite{cmu_xctf}. }
\label{fig:experimental_setup}
}
\end{figure*}

\subsection{Experimental design and setup}
\label{subsec:exp}
We first present the experimental setup of this study (Fig.~\ref{fig:experimental_setup}). A series of bare-plate LPBF experiments were conducted to fabricate Ti-6Al-4V (Ti-64) specimens on an EOS M290 system under controlled conditions ($1064~\mathrm{nm}$ Yb-fiber laser, Gaussian profile, $400~\mathrm{W}$ max power, $100~\mu\mathrm{m}$ spot, Argon shielding). The chemical composition of the Ti-64 alloy, supplied by Allegheny Technologies Incorporated, Pittsburgh, PA, is summarized in Tab.~\ref{tab:powder_chemistry}.

\begin{table}[!ht]
    \centering{
    \caption{Ti-64 chemical composition in weight percentage ($\text{wt}\%$). }
    \vspace{-2mm}
    \resizebox{\linewidth}{!}{
    \begin{tabular}{c|c|c|c|c|c|c|c|c|c}
        \toprule
        \textbf{Element} & Ti & Al & V & C & Fe & H & N & O & Y \\
        \midrule
        \textbf{wt\%} & $\mathrm{Bal.}$ & $6.0300$ & $4.0400$ & $0.0080$ & $0.2000$ & $0.0007$ & $0.0100$ & $0.1520$ & $<0.0009$ \\
        \bottomrule
    \end{tabular}}
    \label{tab:powder_chemistry}}
\end{table}

As shown in Fig.~\ref{fig:experimental_setup}(a), the \textit{in~situ} process monitoring system integrated two primary sensing modalities: a PCB Piezotronics HT378A06 microphone system for capturing airborne AE signals up to $50~\mathrm{kHz}$ in bandwidth, and a Thorlabs PDA10CS2 (A10) photodiode (PHO) sensitive to the fusion laser wavelength of $1064~\mathrm{nm}$. The AE and PHO sensors were interfaced with a National Instruments CompactDAQ (cDAQ) system using an NI9232 data acquisition module. Both channels were sampled at a synchronized rate of 100~$\mathrm{kHz}$, with data collection being initiated concurrently via a shared software triggering mechanism to ensure inter-modal temporal alignment. The AE sensor was pre-mounted inside the build chamber approximately 50~$\mathrm{mm}$ above the build plate, with its position fixed throughout the entire build to ensure consistency in signal acquisition. The PHO sensor was mounted inside the build chamber above the build plate and oriented at a fixed angle toward the fusion zone. In this study, the PHO sensor was employed to detect fluctuations in laser reflections from the fusion zone, enabling real-time identification of the laser on/off status. Although the incident angle of the PHO sensor may influence the signal intensity, its effect on the salient laser reflections was considered negligible for this study~\cite{ren2023machine}.

\begin{figure}[!ht]
\center{\includegraphics[width=\linewidth]
{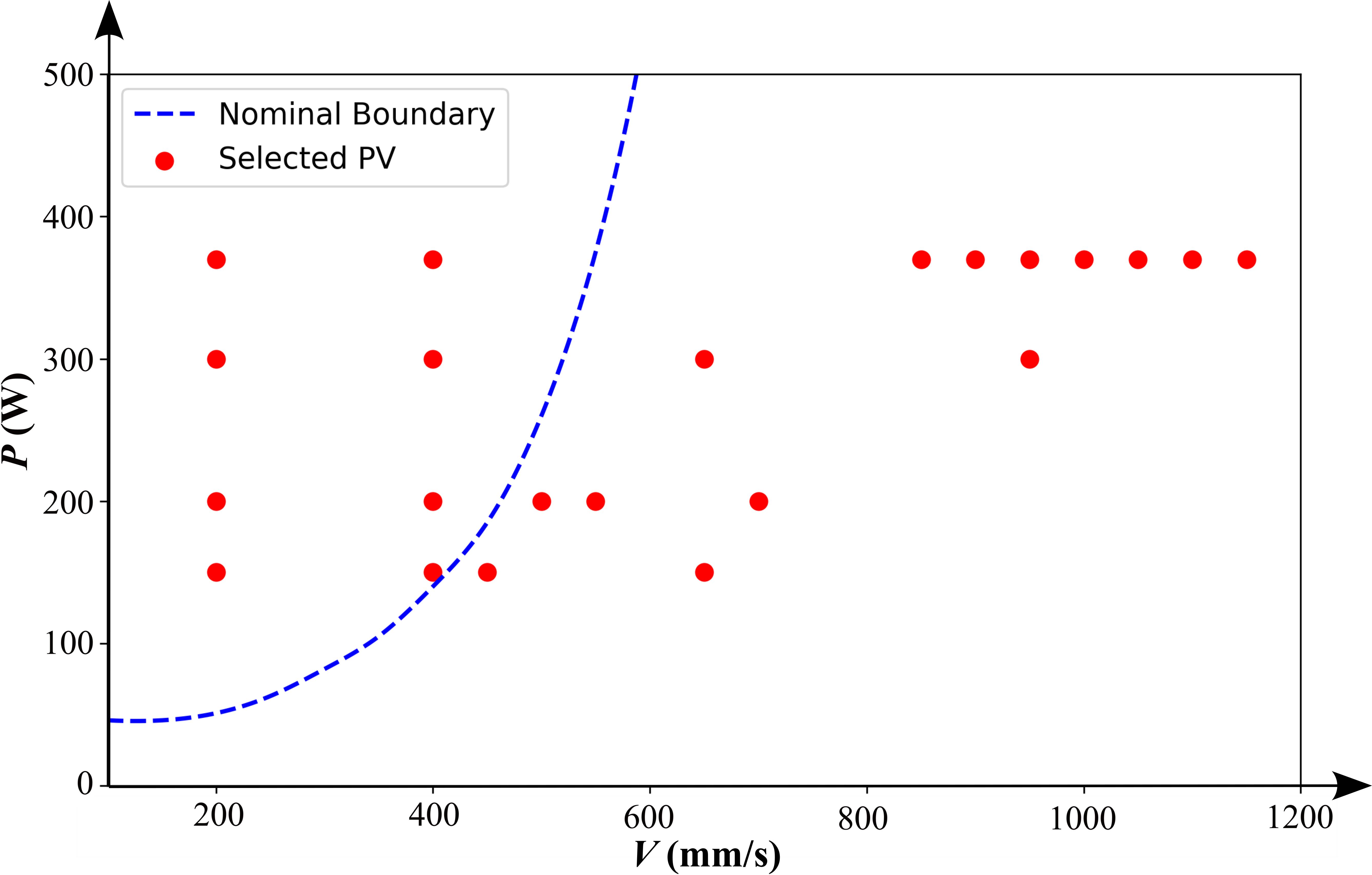}
\caption{Red points represent the selected $P\mathrm{-}V$ points used for the LPBF bare-plate experiments. The blue dashed line is a nominal keyhole-process-window boundary reported by Zhao et al.~\cite{zhao2020critical}. }
\label{fig:PV_design}
}
\end{figure}

The selected $P\mathrm{-}V$ parameters for specimen fabrication (Fig.~\ref{fig:PV_design}) spanned regions within or near the KH and process window regimes, as defined by Zhao et al.~\cite{zhao2020critical}, generating data points from both KH-porous and porosity-free specimens. To study the influence of geometry and scanning strategy on KH porosity, two types of specimens were fabricated for each selected $P\mathrm{-}V$ (shown in Fig.~\ref{fig:experimental_setup}(b)):
\begin{itemize}
    \item \textbf{Single-bead specimens (SBD)}: four $6~\mathrm{mm}$ scan lines spaced $1.5~\mathrm{mm}$ apart to minimize thermal interaction. 
    \item \textbf{Pad specimens (PAD)}: $6\times 6~\mathrm{mm}^2$ rasterized squares with varied scan orientations and hatch spacings between $130~\mu\mathrm{m}$ to $213~\mu\mathrm{m}$ to assess the effects of bead overlap.
\end{itemize}

After printing, the as-built SBDs and PADs were scanned by XCT using a Zeiss Xradia CrystalCT system ($2.96~\mu\text{m}$ voxel size, HE2 filter, $2.35~\mathrm{s}$ exposure, $160~\mathrm{kV}$, $10~\mathrm{W}$). Beam shift and beam hardening parameters were manually adjusted during reconstruction. Porosity segmentation and analysis were performed in Dragonfly~\cite{dragonfly}, yielding detailed porosity characteristics---including size, location, morphology, etc.---which were exported for subsequent training and evaluation of the ML model. (see Fig.~\ref{fig:experimental_setup}(c)).

\subsection{Data processing and cross-modality registration}
\label{subsec:dat}
A multimodal dataset was obtained from the experiments, including airborne AE data, PHO laser reflection data, and XCT porosity data in their respective raw formats. We now introduce the processing and calibration of the dataset for different modalities. 

\subsubsection{Time-series modalities}
\label{subsubsec:time_series}
AE and PHO sensors returned time-resolved signals capturing distinct physical phenomena during the LPBF process. The two streams of signals were sampled at the same frequency, enabling synchronized data acquisition and the construction of a common timeline. 

\textbf{Airborne AE} signals were collected via the ultrasonic microphone system installed inside the build chamber. The sensor captures pressure variations at a fixed spatial position above the build plate, converting them into time-series voltage signals. Sampling at $100~\mathrm{kHz}$ provides sufficient temporal resolution to capture the transient acoustic events associated with laser-material interactions, while also satisfying the Nyquist–Shannon sampling criterion to avoid aliasing. The acoustic latency, corresponding to the time required for the airborne acoustic wave to propagate from the laser fusion zone to the signal reception point, was calculated to be $0.72~\mathrm{ms}$. This delay is non-negligible, given the high temporal resolution required for our AE-driven porosity inference. To account for this latency and to ensure accurate temporal alignment with the PHO laser reflection data, the AE time-series signals were uniformly advanced by $72$ data points for all samples, corresponding to the aforementioned acoustic latency at its sampling frequency.

\textbf{PHO laser reflection} signals were acquired using the PHO sensing system, which consisted of a PHO sensor equipped with a $1064~\mathrm{nm}$ band-pass filter. Operating on a principle analogous to that of the airborne AE sensor, the PHO system converts variations in laser reflection intensity into time-series voltage signals. Consistent with AE, the PHO signals were sampled at a frequency of $100~\mathrm{kHz}$. The primary use of the PHO data in this study was to detect laser on/off events by exploiting the distinct contrast in laser reflection intensity between the on and off states, enabling the segmentation of individual scanlines and layers from the very long, continuous stream of experimental data. Consequently, the PHO signal only served as a temporal marker to help segment the AE data into its salient clips. 

\begin{figure}[!ht]
\center{\includegraphics[width=0.8\linewidth]
{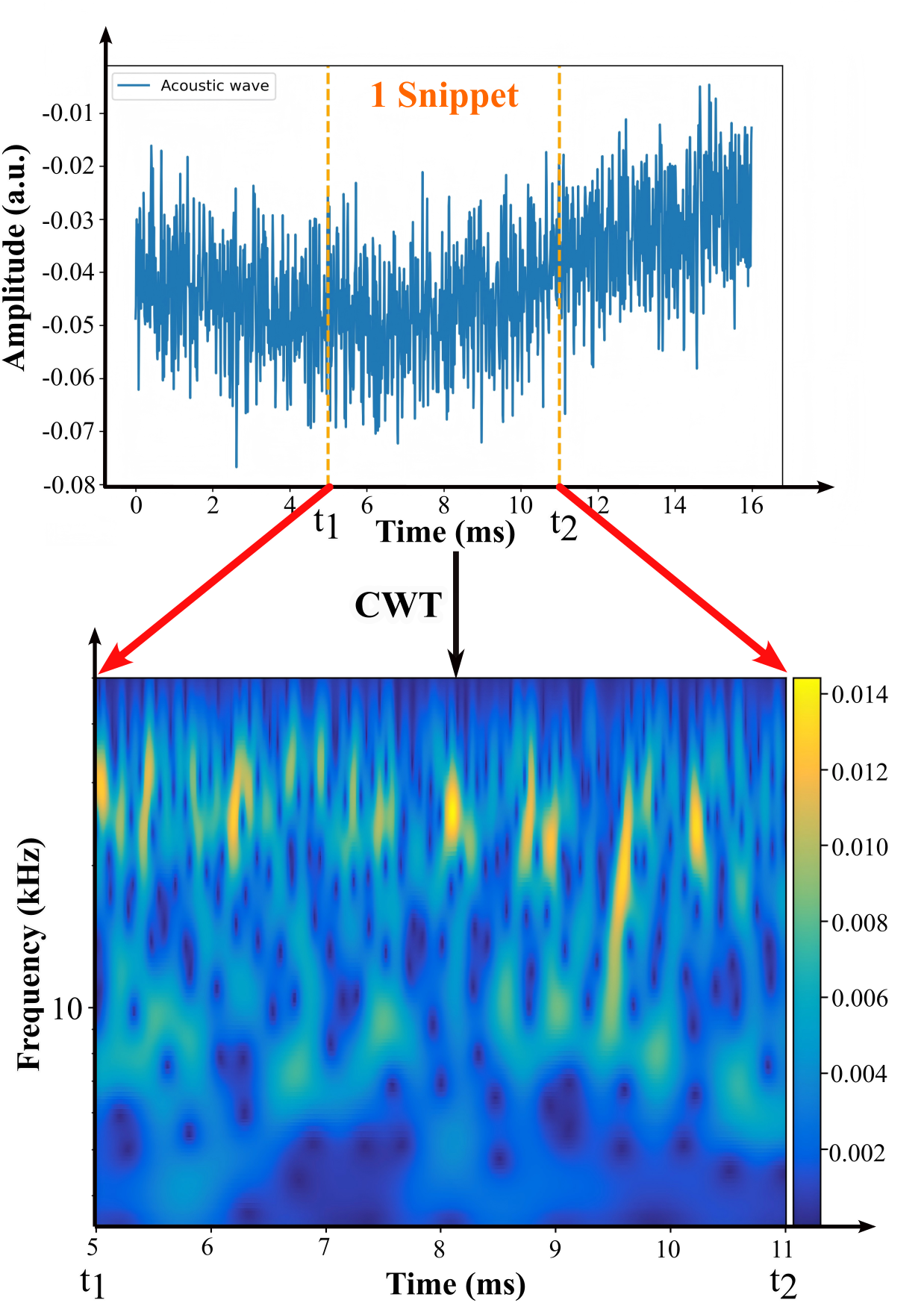}
\caption{Temporal clipping of an AE snippet and the corresponding AE scalogram snippet after CWT processing. }
\label{fig:cwt}
}
\end{figure}

\textbf{Clipping and spectrotemporal analysis} was performed on the AE signals to obtain the time-resolved frequency characteristics after pre-processing and temporal alignment of the time-series data. As illustrated in Fig.~\ref{fig:cwt}, each AE signal was first segmented into short temporal snippets spanning a few milliseconds, whose length was varied as described in subsequent sections. Then, each AE segment was transformed into the time–frequency ($t\mathrm{-}f$) space via continuous wavelet transform (CWT), implemented using the \texttt{ssqueezepy} package in \texttt{Python}~\cite{qin2021fcanet, OverLordGoldDragon2020ssqueezepy}. The \texttt{Morlet} wavelet was selected as the mother wavelet due to its favorable $t\mathrm{-}f$ localization properties. This wavelet was systematically scaled and phase-shifted to generate a family of wavelets, enabling the decomposition of AE signals and the computation of inner product similarity between wavelet and acoustic functions across varying timestamps and frequency bands. This process produced a high-resolution representation of the acoustic energy distribution, called a scalogram, in the $t\mathrm{-}f$ domain, allowing a detailed capture of transient acoustic characteristics associated with KH porosity formation. We note that PHO laser reflection data was utilized solely for data synchronization as described above. As such, no spectral analysis was performed on the PHO data. 

\subsubsection{XCT for porosity quantification}
\label{subsubsec:XCT}
XCT scanning was performed on the as-built SBDs and PADs to capture the internal KH porosity with high spatial resolution ($2.96~\mu\mathrm{m}$). Following 3-D volumetric reconstruction, Dragonfly was used for pore segmentation and quantitative analysis. The segmentation process involved: (1) global intensity thresholding to isolate low-density regions, (2) connected component labeling to distinguish discrete pores, and (3) morphological filtering to remove artifacts and ensure segmentation fidelity. Manual verification was conducted to correct over- or under-segmented regions, particularly in areas near the fusion boundaries. After pore segmentation, the software automatically extracted multiple porosity characteristics for each pore, including the voxel-based size, the aspect ratio, the Feret diameter, and others. 

\textbf{KH porosity data extraction and filtering} were subsequently implemented to export the porosity datasets as structured spreadsheets, each of which corresponded to a single laser scanline. Each row in a spreadsheet corresponded to an individual pore or artifact. The columns encoded various geometric and spatial descriptors available through Dragonfly’s analysis suite. Critically, we extracted the horizontal distance between each pore center and the start of the scanline, as this distance (together with the knowledge of the laser speed) facilitates a spatiotemporal registration of the pores with the AE data at sub-scanline resolution. Furthermore, post-processing filters were applied to the XCT-derived pore data to eliminate spurious detections. Notably, pores with a voxel count below $4$ or a minimum Feret diameter less than $5.92~\mu\mathrm{m}$ were excluded, as these fell below the empirically determined thresholds to distinguish KH pores from segmentation artifacts. Additionally, pores located within $0.1~\mathrm{mm}$ or at the start and end of each scanline were removed to mitigate boundary effects introduced by laser deceleration and acceleration during turnarounds, which were not expected to produce reliable AE signatures captured by our regression model. The filtering criteria discussed above ensured that only physically meaningful AE-detectable KH pores were retained for the subsequent model development.

\begin{figure}[!ht]
\center{\includegraphics[width=\linewidth]
{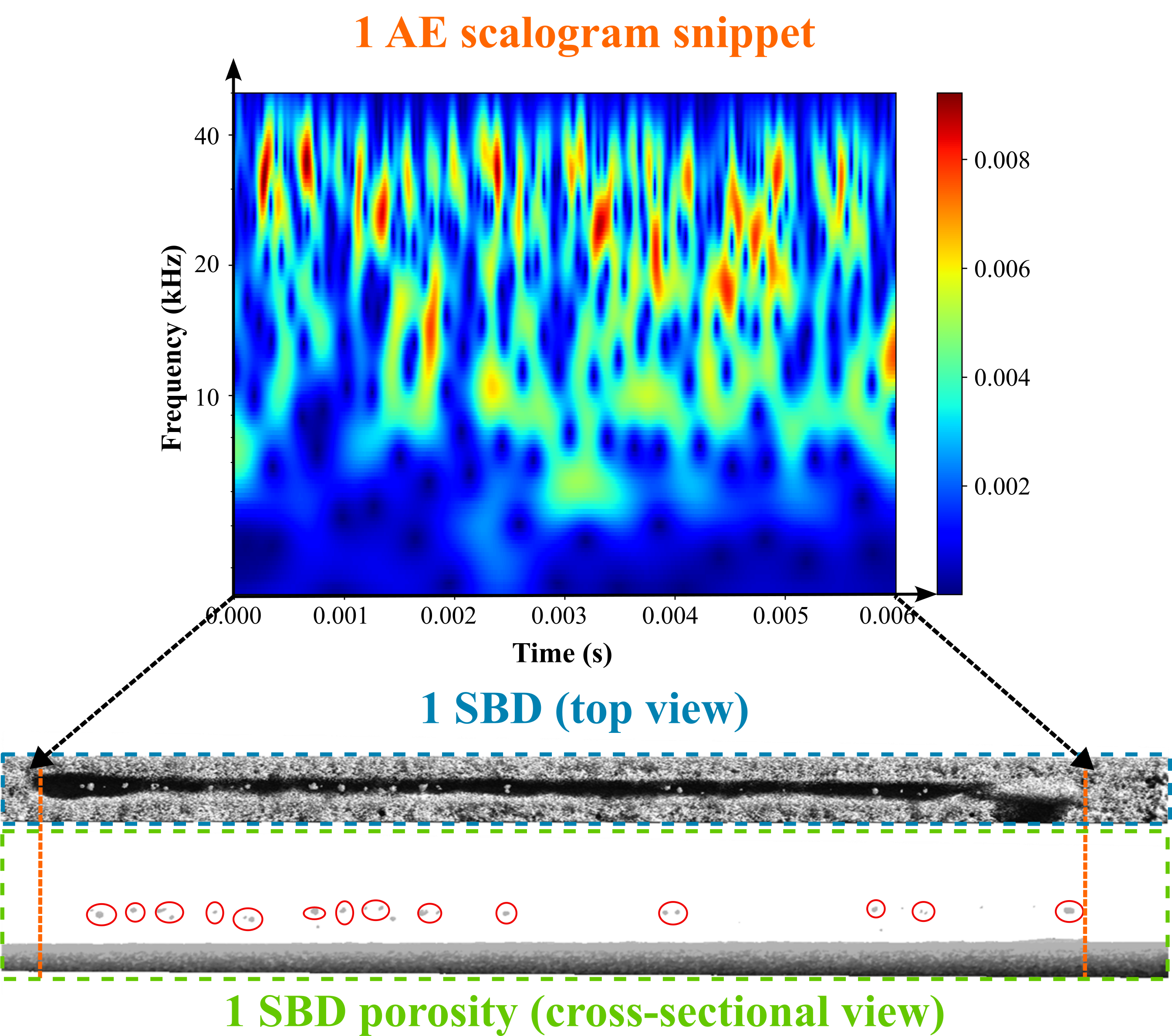}
\caption{Spatiotemporal registration between AE scalogram snippets and the corresponding XCT porosity image segment. The example shown illustrates a full SBD, though in practice, the aligned segment may span less than a full scanline depending on the window length used. }
\label{fig:spatiotemporal}
}
\end{figure}

\textbf{Spatiotemporal data registration} between the XCT-based porosity data (spatial) and the AE signals (temporal) is a fundaamental component for accurate pore quantification within our proposed framework. As depicted in Fig.~\ref{fig:spatiotemporal}, the start and stop timestamps were calculated relative to the onset of its associated scanline for each AE snippet. An accurate spatiotemporal alignment requires determining: (a) the scanline/spreadsheet to which the AE snippet belongs, (b) the spatial location of the starting point of that scanline, and (c) the laser scan speed~($V$). These parameters were available through the bead morphological patterns, process parameter profiles, and active scan profiles identified from the PHO laser reflection data. Using this information, the spatial segment of the scanline traversed during the AE snippet was located, enabling the extraction of all affiliated pores and artifacts from the XCT-generated data spreadsheets. 

\textbf{KH pore drift accommodation} was carried out to enhance the robustness of the AE snippet data labeling. Due to melt pool convection and turbulence, KH pores generated during the process may drift rearward within the molten region over a non-negligible distance on the order of tens to hundreds of microns within a millisecond timescale~\cite{pang2014quantitative}. In this study, it was challenging to directly characterize the drifting displacement of individual KH pores, owing to the lack of real-time X-ray synchrotron imaging and \textit{in~situ} pore tracking during fabrication. To account for this limitation and reduce the potential misattribution of pores to incorrect AE snippets, a fixed pore-drift tolerance of 50~$\mu\mathrm{m}$ was appended to the end of the spatial window corresponding to each AE snippet, irrespective of $V$. Although improving robustness against miscounting, this approach may introduce the risk that overly short AE snippets may yield a disproportionately large fraction of drifted pores within the target scanline segments, and consequently compromise alignment fidelity. To mitigate this risk, a lower bound of $4~\mathrm{ms}$ duration was imposed on the time window for the AE snippet clipping throughout data processing in Sec.~\ref{subsubsec:time_series}.

\subsection{Data-driven ML model}
\label{subsec:ML}
We then describe our ML approach for KH porosity prediction. Dataset curation, model training, and validation were all implemented in Python~\cite{van1995python, harris2020array, 2020SciPy-NMeth, mckinney-proc-scipy-2010, Hunter:2007, clark2015pillow}. Training and evaluation were performed using PyTorch 1.13.1~\cite{NEURIPS2019_9015} on a workstation equipped with an Intel(R) Core(TM) i7-10700K CPU @ 3.80GHz, an NVIDIA GeForce RTX 3070 GPU with 8GB of VRAM, and CUDA 12.0.

\subsubsection{ML dataset establishment}
\label{subsubsec:ML_dataset}
As outlined in Sec.~\ref{subsec:dat}, a multimodal dataset from LPBF Ti-64 SBD and PAD bare-plate printings was curated with rigorous spatiotemporal intermodal data synchronization. To support streamlined data labeling for ML purposes, we implemented a unified data indexing system that enables the retrieval of all relevant data. This includes the AE snippet, PHO laser reflections, and scanline segments derived from XCT that contain valid keyhole porosities, all packaged together based on the timestamp of the snippet. 

\textbf{Output label---$\texttt{KHLineNum}$}, a quantifiable measure derived from observed KH porosities, was used as the ground truth output label for supervised learning. Although previous work has frequently relied on macroscale measures such as the volumetric porosity fraction or the total KH pore count to quantify the severity of KH~\cite{wang2022analytical}, these aggregate indicators lack spatial specificity regarding the location of individual KH pores. In addition, Tempelman et al.~\cite{tempelman2022detection, tempelman2024uncovering} and Ren et al.~\cite{ren2024sub} have demonstrated the link between the acoustic features and resulting KH pore formation events. However, the use of acoustics for a fine-grained quantification of pore generation remains an unmet challenge. To this end, we create a new KH porosity metric by characterizing local porosity generation as the number of KH pores generated per unit length along the laser scanning trajectory. We denote this as the scalar-valued $\texttt{KHLineNum}$~($\mathrm{\mu m}^{-1}$), expressed formally as: 
\begin{equation}
\label{eqn:KHLineNum}
    \mathrm{\texttt{KHLineNum}}=\frac{N_{\mathrm{pores}}}{L_{\mathrm{travel}}}
\end{equation}

\noindent where $N_{\text{pores}}$ denotes the number of KH pores and $L_{\text{travel}}$ denotes the distance traversed by the laser within a given snippet. Eq.~\eqref{eqn:KHLineNum} defines a scalar metric that quantifies the spatial rate of KH porosity generation at a given moment. As demonstrated in Sec.~\ref{subsec:qualitative_analysis}, $\texttt{KHLineNum}$ is promising for AE-based KH porosity inference, enabling higher spatiotemporal resolution in characterizing KH porosity distributions. 

\textbf{Input data} was constructed as an image-derived dual-channel 3-D data matrix. The AE scalogram (see Sec.~\ref{subsec:dat}) was used as the first channel of the input tensor. Each AE scalogram image was bilinearly interpolated, rescaled, and mapped to a colormap image, which was then converted into a single-channel grayscale image of size~$224\times224$. In this image, the horizontal axis represents time, and the vertical axis spans frequencies from $5~\mathrm{kHz}$~to~$50~\mathrm{kHz}$ on a logarithmic scale. Pixel intensities reflect the CWT amplitude, indicating localized energy in the $t\mathrm{-}f$ domain. $V$ was incorporated as the second input channel. Each scalar $V$ value was normalized to the range $\left[0,~255\right]$ and tiled into a matrix matching the scalogram’s resolution. The scalogram and $V$ matrix were then stacked to produce a $224\times224\times2$ input tensor. As discussed in Sec.~\ref{subsec:qualitative_analysis}, this combination of AE scalograms and $V$ yielded the best predictive performance for $\texttt{KHLineNum}$, a result supported by iterative ML experiments conducted after the initial qualitative assessment.

\textbf{Dataset partitioning} was conducted to curate two datasets from the SBD and PAD builds, respectively. For training, a subset of AE snippets from the SBD dataset was used. The data was shuffled at the $P\mathrm{-}V$ parameter level, with $80\%$ of the combinations used for training and $20\%$ held out for validation. The PAD dataset, which includes all combinations of $P$, $V$, and $H$, was reserved entirely for additional validation. To ensure generalizability and prevent data leakage, strict partitioning between training and validation sets was enforced. Multiple versions of the dataset were created by varying AE snippet durations from $4~\mathrm{ms}$ to $10~\mathrm{ms}$. These were blended in training, while maintaining consistent $P\mathrm{-}V$ pairings. The final dataset contained approximately $3,200$ labeled snippets, encompassing all combinations of $P\mathrm{-}V$ settings as well as various time durations.

\subsubsection{Architecture of the ML model}
\label{subsubsec:ML_archi}
\begin{figure}[!ht]
\center{\includegraphics[width=\linewidth]
{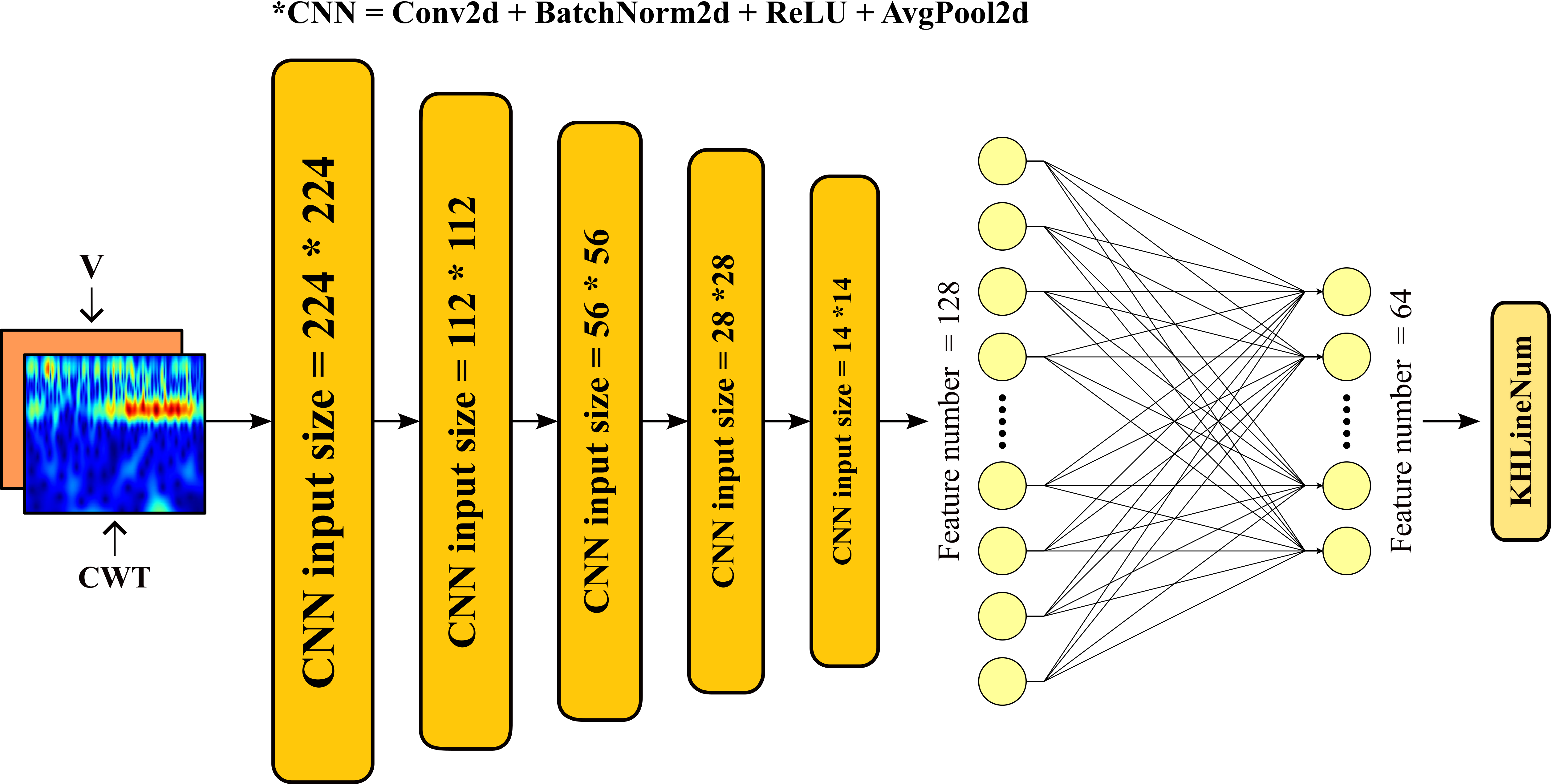}
\caption{A tractable ML architecture for AE-driven $\texttt{KHLineNum}$ predictions, consisting of five convolutional blocks and two fully connected layers. The model maps image-like input data to the targeted KH porosity scalar metric. }
\label{fig:ML_pipeline}
}
\end{figure}

Each input tensor was processed by a CNN followed by a fully connected multi-layer perception (MLP) to predict the corresponding $\texttt{KHLineNum}$. The CNN comprised five $3\times 3$ convolutional layers (with stride 1, padding 1) with ReLU activations and average pooling after each layer, as shown in Fig.~\ref{fig:ML_pipeline}. The input tensor was progressively downsampled and encoded into a $7\times 7\times 128$ latent representation by the convolutional backbone, with feature depth increasing at each layer. This tensor was flattened and passed to an MLP with two fully connected layers ($128$ and $64$ units) for scalar prediction.

\subsubsection{Training and validation details}
\label{subsubsec:ML_train_val}
The model was trained using a batch size of $64$ for $150$~epochs, with an initial learning rate (LR) of~$1\times 10^{-4}$. The~\texttt{Adam} optimizer and an LR scheduler were used to train the model and reduce LR after stagnation of the loss value, respectively. The loss function was mean squared error (MSE) between the predicted and ground truth $\texttt{KHLineNum}$ values, which were logarithmically transformed to stabilize training and enhance contrast in label distribution. The performance of the model was evaluated using the coefficient of determination ($R^2$) in both the SBD and PAD validation datasets. For PAD evaluations, the predicted outputs were exponentiated and summed to reconstruct total pore counts, enabling comparison against ground truth KH porosity under realistic build conditions.

\section{Results and analysis}
\label{sec:results}
\begin{figure*}[!ht]
\center{\includegraphics[width=\linewidth]
{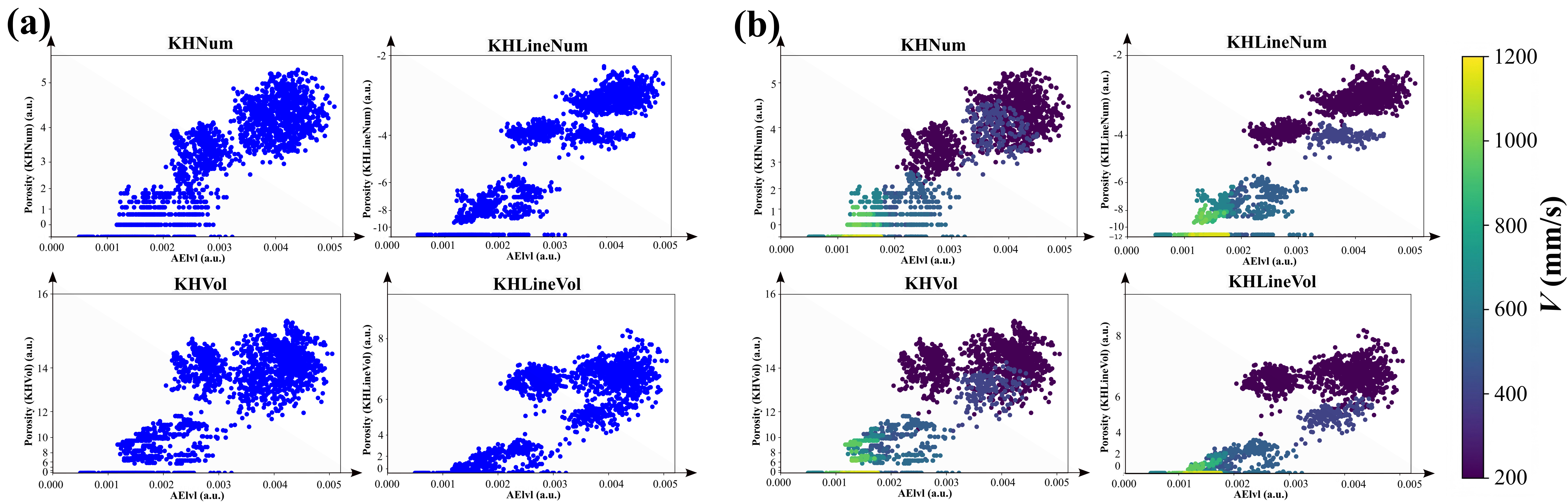}
\caption{Qualitative correlation analysis between $\texttt{AElvl}$ and KH-porosity-related measures across snippets of varying durations. $\texttt{AElvl}$~vs.~KH porosity scatter plots that are (a)~monocolor and (b)~$V$-annotated are showcased in respective 4-plots. }
\label{fig:qual}
}
\end{figure*}

\subsection{Qualitative analysis on AE to KH-porosity correlations}
\label{subsec:qualitative_analysis}
We began by qualitatively assessing the relationship between AE scalogram characteristics and KH porosity. Each AE scalogram was reduced to a single scalar value, $\texttt{AElvl}$, defined as the average CWT intensity and serving as a coarse representation of the scalogram tensor. For each snippet, $\texttt{AElvl}$ was paired with one of four KH porosity metrics: total volume ($\texttt{KHVol}$), total count ($\texttt{KHNum}$), line-density of volume ($\texttt{KHLineVol}$), and line-density of count (\textit{i.e.}, $\texttt{KHLineNum}$). These relationships are shown as 2-D scatter plots in Fig.~\ref{fig:qual}(a), with $\texttt{AElvl}$ on the $x$-axis and the corresponding porosity metric on the $y$-axis. Vertical axis values were logarithmically transformed for contrast, with zeros replaced by a small constant to avoid undefined \texttt{log} operations.

Observing the results in Fig.~\ref{fig:qual}(a), we draw the following three conclusions:
\begin{enumerate}
    \item A higher $\texttt{AElvl}$ generally corresponds to a higher KH porosity, regardless of the specific KH porosity metric.
    \item For every pairwise plot, a given $\texttt{AElvl}$ can map to a wide range of porosity, indicating that even a full 2-D scalogram, though richer than a single scalar, may still lack the resolution needed for precise quantification of KH porosity.
    \item Of the four porosity metrics, $\texttt{KHLineNum}$ yields the most distinct clusters with minimal overlap, outperforming $\texttt{KHNum}$, $\texttt{KHVol}$, and $\texttt{KHLineVol}$. This corroborates our claim in Sec.~\ref{subsubsec:ML_dataset} that $\texttt{KHLineNum}$ is the most discriminative and learnable target for the present feature space.
\end{enumerate}

These results reinforce the need to augment the AE scalogram with complementary features to improve predictive fidelity. Accordingly, Fig.~\ref{fig:qual}(b) replots the data, coloring each point by~$V$~($\mathrm{mm/s}$) to visualize how process parameters modulate the $\texttt{AElvl}$-porosity relationship. The colored 4-plot shows that $\texttt{KHLineNum}$ remains the most separable porosity metric when the characteristics of the AE scalogram are combined with the $V$ annotation. These findings suggest that within the $\texttt{AElvl}\mathrm{-}V$ feature space, a distinct and consistent level of $\texttt{KHLineNum}$ can potentially be inferred for almost any given point, thereby corroborating our claim in Sec.~\ref{subsubsec:ML_dataset}. However, we also noticed that $V$ alone is weakly discriminative, since points with similar $V$ span a wide range of KH porosity levels, highlighting the importance of maintaining the temporally resolved AE features. We therefore hypothesized that leveraging a richer $t\mathrm{-}f$ representation of AE and the $V$ parameter should outperform models based solely on the scalar $\texttt{AElvl}$. 

Lastly, we also evaluated the impact of appending $P$ to the input space, hypothesizing that it might improve performance as a proxy for fusion energy density. Contrary to expectations, incorporating $P$ consistently degraded model performance across repeated trials. Further details of these findings are provided in~\ref{appendix:qual_addi}.

\subsection{Quantification of KH porosities on SBD dataset}
\label{subsec:pred_acc}
\begin{figure}[!hbt]
\center{\includegraphics[width=\linewidth]
{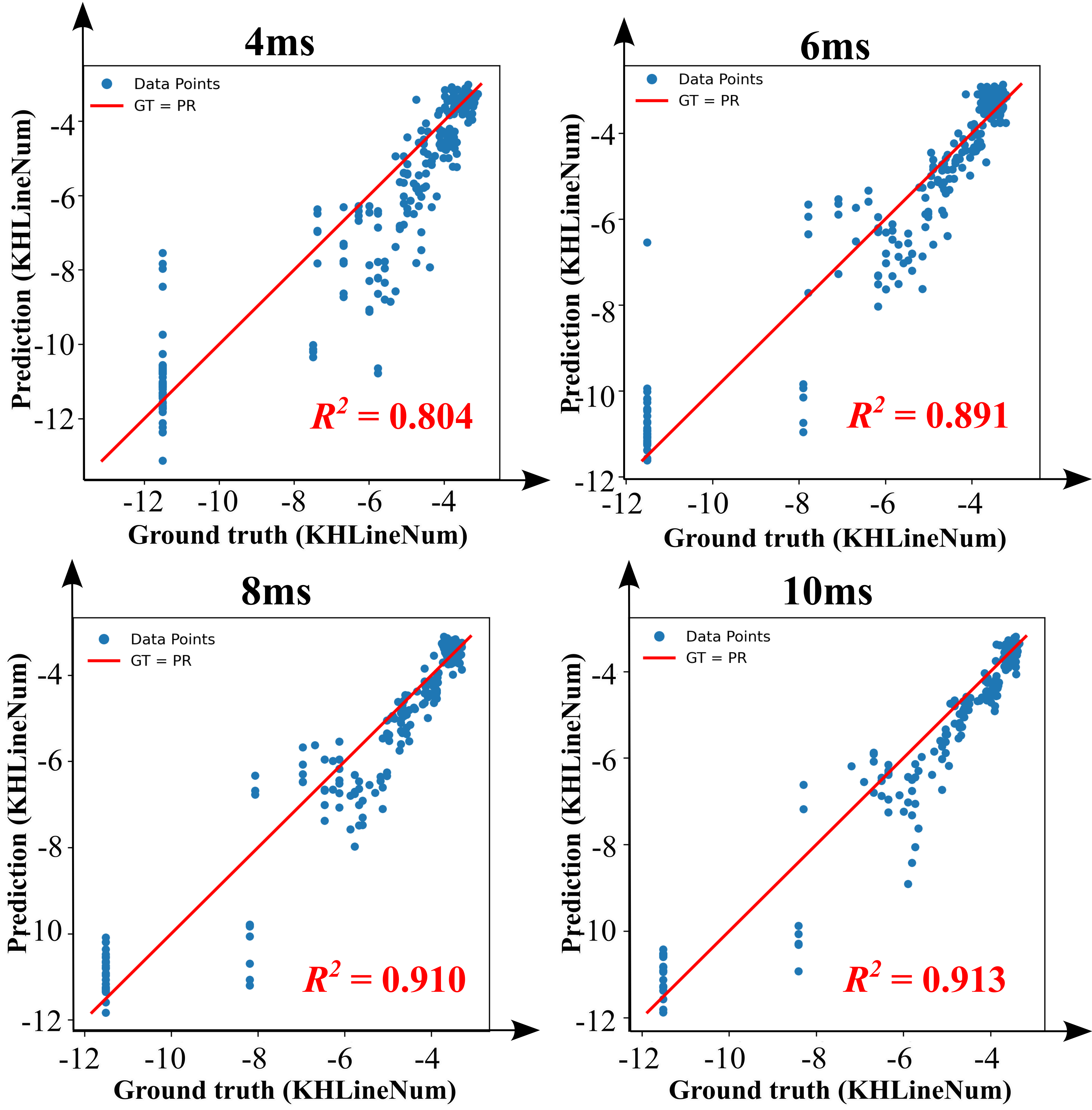}
\caption{$R^2$ plots of $\texttt{KHLineNum}$ predictions across varying AE snippet durations for SBDs. A gradual decline in prediction accuracy is observed as snippet duration decreases. }
\label{fig:r2_SBD}
}
\end{figure}

After training, the model was evaluated on the SBD validation subset. The random shuffle described in Sec.~\ref{subsubsec:ML_dataset} preserved a consistent $\texttt{AElvl}\mathrm{-}V$ distribution across training and validation splits. We further grouped the validation data by durations to gauge the effect of snippet length. Figure~\ref{fig:r2_SBD} shows the ground truth $\texttt{KHLineNum}$ ($x$-axis) against the predictions ($y$-axis) for each duration of the snippet, with the corresponding $R^2$ values shown in each subplot.

From Fig.~\ref{fig:r2_SBD}, we draw the following three conclusions: 
\begin{enumerate}
    \item $\texttt{KHLineNum}$ can be inferred from AE scalograms and $V$ with $R^2>0.8$.  
    \item Prediction performance depends on the duration of the snippet: $R^2$ falls for shorter snippets, possibly reflecting the loss of temporal context. 
    \item For instances where the ground truth values of $\texttt{KHLineNum}$ are $0$, a vertically aligned cluster of inaccurate predictions appears across all plots in Fig.~\ref{fig:r2_SBD}, indicating the model’s limited efficacy and generalizability under conditions of lower fusion energy density.    
\end{enumerate}

The reduced prediction accuracy under low fusion energy density conditions may be attributed to three factors: (1) the model’s tendency to misinterpret random noise as signal, (2) the transient formation and rapid healing of KH pores within very short time intervals, and (3) reduced vaporization at low fusion energy density, which decreases AE generation. In contrast, for medium- and high-porosity samples, predictions remain stable and accurate, highlighting the robustness of AE-based $\texttt{KHLineNum}$ inference when porosity features are more prominent. Moreover, Fig.~\ref{fig:r2_SBD} indicates that a $6~\mathrm{to}~8~\mathrm{ms}$ snippet, given the AE sampling rate of $100~\mathrm{kHz}$, provides an optimal trade-off between prediction accuracy and temporal resolution. Consequently, extending the temporal context, if feasible, may further enhance the quantification performance. 

\begin{figure}[!ht]
\center{\includegraphics[width=\linewidth]
{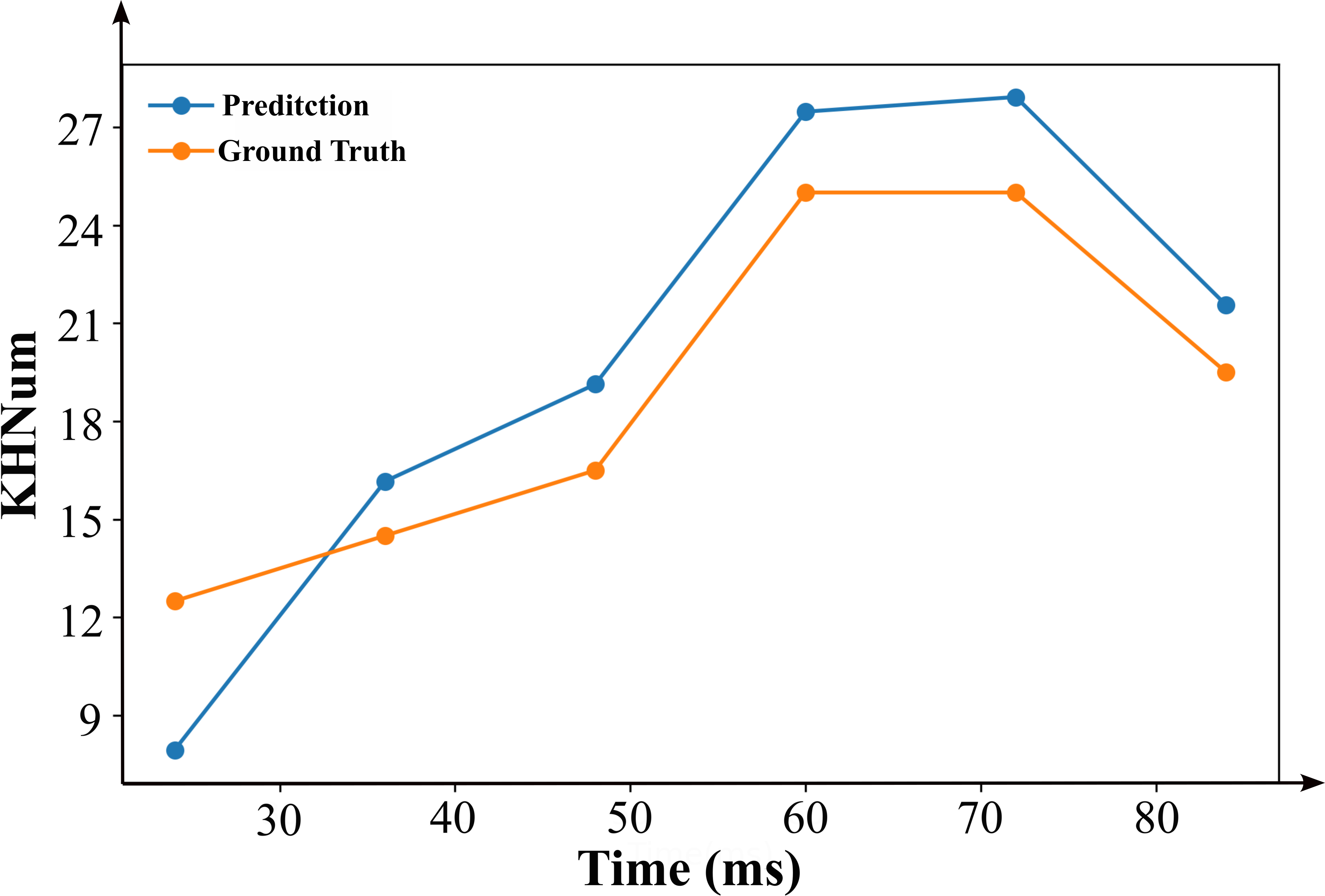}
\caption{Temporal tracking of KH porosity generation for an SBD from the test dataset using a non-overlapped $12~\mathrm{ms}$ time window. }
\label{fig:SBD_ts}
}
\end{figure}

To demonstrate the temporal tracking performance, Fig.~\ref{fig:SBD_ts} presents a comparison between the predicted and ground truth $\texttt{KHNum}$ for a representative SBD build exhibiting medium to high porosity. A non-overlapping $12~\mathrm{ms}$ sliding window was employed to traverse the entire build, generating consecutive predictions of $\texttt{KHLineNum}$, which were subsequently scaled to obtain $\texttt{KHNum}$. Despite the slight overestimation observed, the predicted curve aligns well with the measured trend, supporting the effectiveness of the framework for potential near-real-time monitoring of short-duration defect events.  

\begin{figure}[!ht]
\center{\includegraphics[width=\linewidth]
{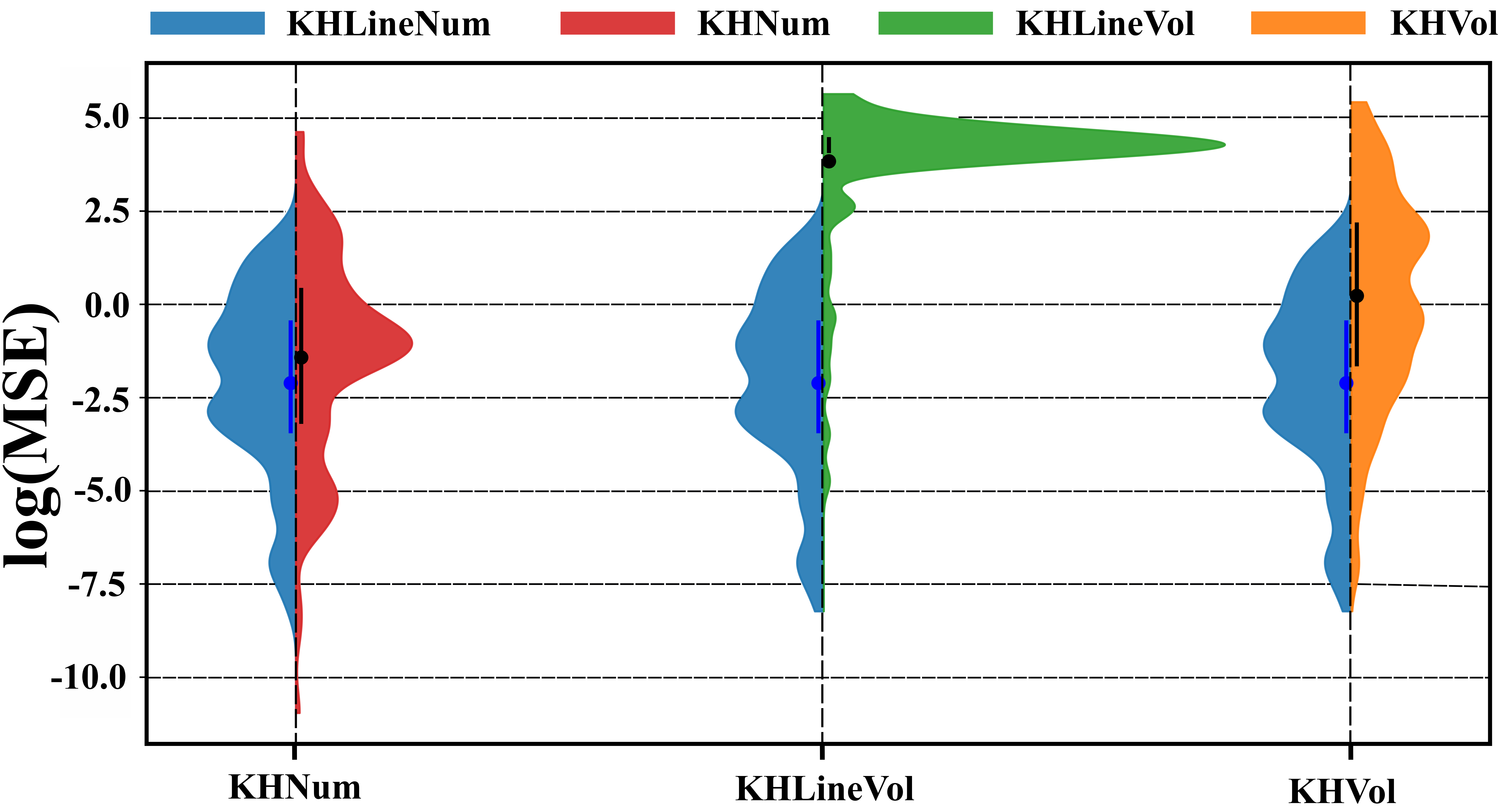}
\caption{Violin plots comparing the log-transformed prediction error ($\log\left(\mathrm{MSE}\right)$) for $\texttt{KHLineNum}$ versus alternative KH porosity measures. Each violin plot represents the distribution of model discrepancies across SBD test samples. Results show that using $\texttt{KHLineNum}$ as the target metric yields the best predictive accuracy. The central dot and vertical line in the middle of each violin denote the median and interquartile range of $\log\left(\mathrm{MSE}\right)$, respectively. }
\label{fig:SBD_violins}
}
\end{figure}

In order to validate the selection of $\texttt{KHLineNum}$ as the target variable and extend the analysis in Sec.~\ref{subsec:qualitative_analysis}, we applied the same ML framework to predict alternative KH porosity metrics using a fixed $6~\mathrm{ms}$ snippet duration. Figure~\ref{fig:SBD_violins} presents the resulting MSEs as violin plots: the left half shows the distribution for $\texttt{KHLineNum}$ (as in Fig.~\ref{fig:r2_SBD}), while the right half displays distributions for other metrics. $\texttt{KHLineNum}$ consistently yields the lowest MSE, confirming it as the most suitable ML target. These results align with prior studies~\cite{tempelman2022detection, tempelman2024uncovering, ren2023machine}, reinforcing the effectiveness of AE-based methods for inferring discrete KH pore formations using a well-defined metric such as $\texttt{KHLineNum}$. 

\subsection{Model generalizability on the PAD dataset}
\label{subsec:gen_PADs}
\begin{figure}[!ht]
\center{\includegraphics[width=\linewidth]
{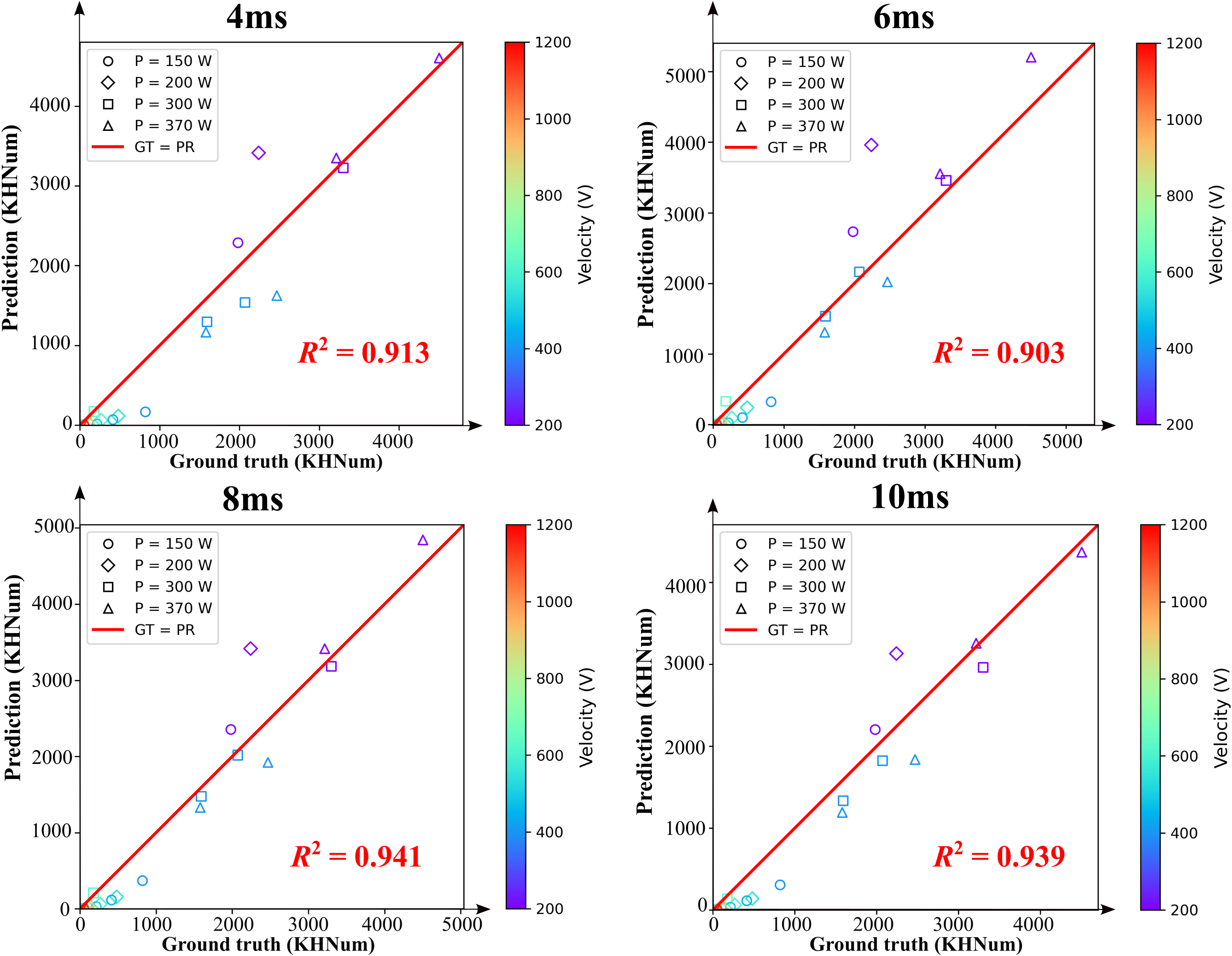}
\caption{$R^2$~plots of $\texttt{KHNum}$ predictions for PAD dataset with different time window durations. Colors and marker types represent~$P$~and~$V$ of the fusion laser, respectively. }
\label{fig:r2_PAD}
}
\end{figure}

To test our method's generalizability over more complicated scanning patterns, we applied the SBD-trained model directly to the PAD dataset without fine-tuning. For each PAD build, AE snippets were extracted with sliding windows of varying length and fed to the model to predict the total $\texttt{KHNum}$ under the PAD layer. From nearly $20$ PADs, we generated $R^2$ plots for each window size, shown in Fig.~\ref{fig:r2_PAD}, demonstrating that the model remained strong, averaging $R^2=0.924$, with performance declining slightly as the window shortened, which echoes the trend in Fig.~\ref{fig:r2_SBD}. This confirms that longer temporal context improves $\texttt{KHLineNum}$ as well as $\texttt{KHNum}$ estimation. Overall, the results show that AE-based models can quantify the KH porosity layer-by-layer in realistic builds, supporting cost-efficient data-driven monitoring in KH-prone regimes. 

As shown in Fig.~\ref{fig:r2_PAD}, the pre-trained model consistently underestimates $\texttt{KHNum}$ for PADs fabricated under low fusion energy density conditions. This again suggests a potential limitation in the model’s generalizability to regimes characterized by sparse KH pore formation, where the acoustic signatures may be weak or ambiguous. Such underestimation suggests that further training or adaptation may be necessary for robust performance in low-defect conditions. 

Following the strategy employed in the PAD validation, we assessed the generalizability of the data-driven model by training additional ML architectures---including ResNets and vision transformers (ViT)---with the same SBD dataset and validating them on the PAD dataset. The corresponding validation $R^2$ values and comparative analyses are presented in~\ref{appendix:ML_models}. 

\subsection{Ablation study and feature importance analysis}
\label{subsec:ablation}
\begin{figure*}[!ht]
\center{\includegraphics[width=\linewidth]
{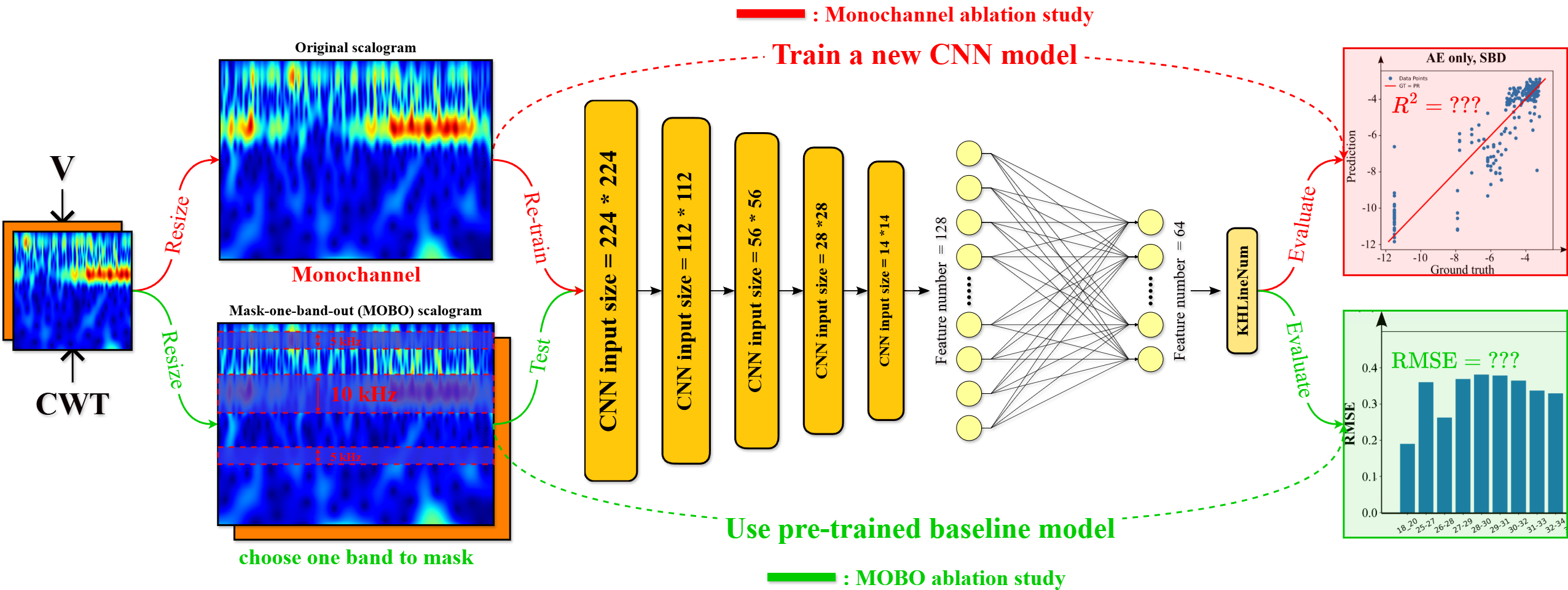}
\caption{Investigation strategy for the ablation studies conducted by altering the model input. The red pathway (top) illustrates the input-channel ablation study, in which new CNN models are trained using the same scheme but with respective monochannel inputs. The green pathway (bottom) represents the frequency-band ablation study, where the baseline model pretrained on SBDs was kept fixed while validating MOBO dual-channel inputs. Eventually, the feature importance of red and green routes was assessed via the comparison of corresponding $R^2$ plots and relative MSE loss values, respectively. }
\label{fig:ablation}
}
\end{figure*}

To better interpret the previous results, we conducted a series of ablation studies to evaluate the relative contributions of different input features. The general investigation pipeline, shown in Fig.~\ref{fig:ablation}, facilitates the analysis focusing on two aspects: (1) the respective importance of the AE scalogram and $V$, and (2) the influence of specific frequency bands within the AE scalogram on the quantification of KH porosity. During model training or validation, we systematically removed selected features and measured the resulting change in accuracy to assess their impact. In the end, we chose to evaluate the prediction accuracy in their corresponding $R^2$ or loss values for either SBDs or PADs and compared them with the results obtained from the previously trained baseline model. 

\subsubsection{AE scalogram~$\mathrm{vs.}$~$V$}
\label{subsubsec:abla_AE_V}
\begin{figure}[!ht]
\center{\includegraphics[width=\linewidth]
{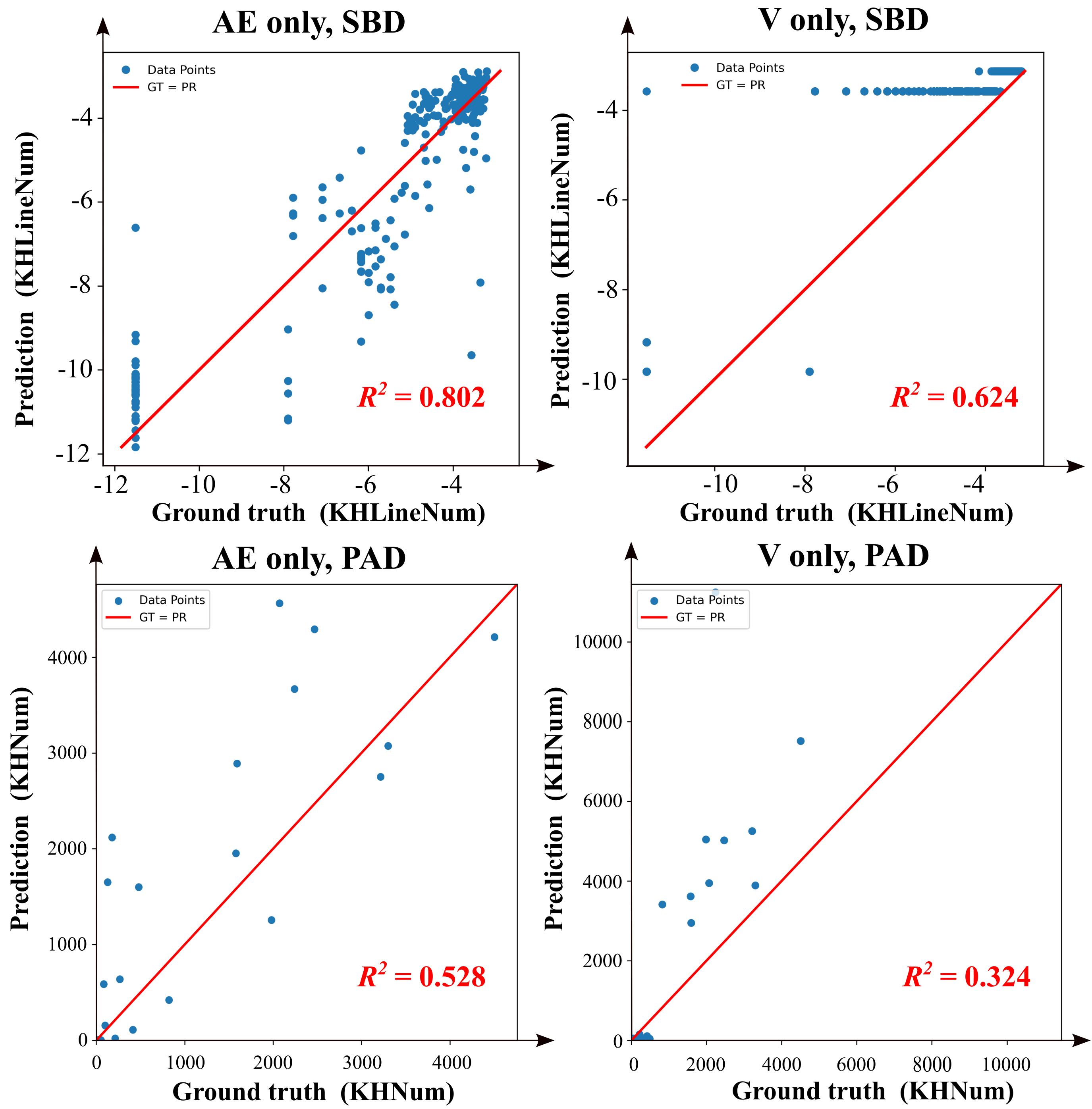}
\caption{The prediction performance of both ablated models on the SBD test dataset (top row) as well as the entire PAD validation datasets (bottom row). One (left column) model received only the AE scalogram as input, while the other (right column) was provided with only the $V$ constant image. The aforementioned optimal time window of $6~\mathrm{ms}$ is used throughout the retraining and the subsequent validation. }
\label{fig:r2_ablation_channel}
}
\end{figure}

We first retrained two separate CNN models using the SBD dataset with either $\texttt{KHLineNum}$ (SBD validation) or $\texttt{KHNum}$ (PAD validation) as the target. Specifically, one of the two models used only the AE scalogram, and the other used only the $V$-constant image. Their performance on the $6~\mathrm{ms}$ SBD and PAD validation sets is shown in Fig.~\ref{fig:r2_ablation_channel}. In all cases, $R^2$ dropped significantly compared to the baseline model evaluated above. This confirms that excluding either the AE scalogram or $V$ degrades the performance of the model and supports our findings in Sec.~\ref{subsec:qualitative_analysis}. 

\begin{figure}[!ht]
\center{\includegraphics[width=\linewidth]
{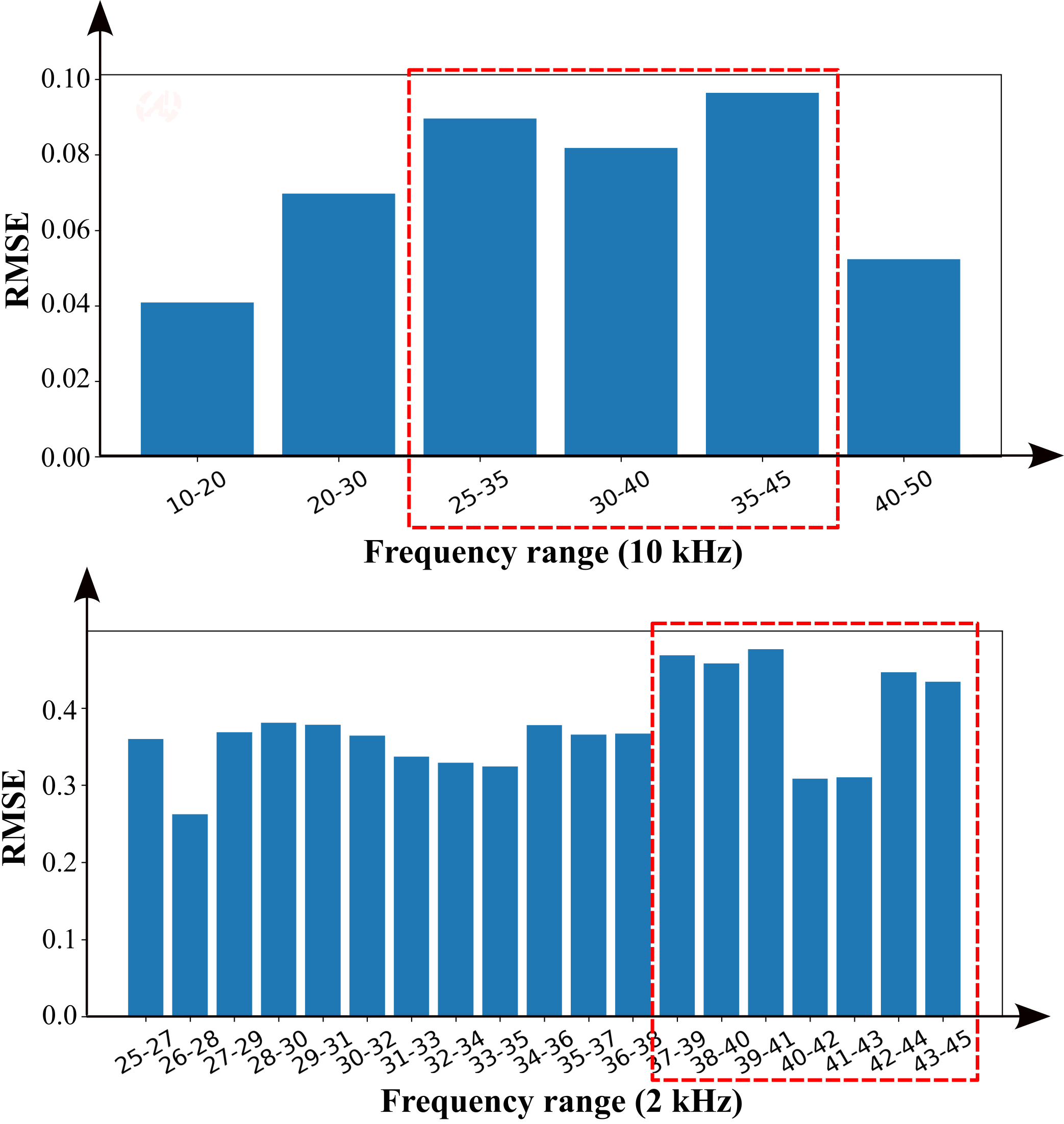}
\caption{RMSE-based feature importance analysis using the fixed pre-trained baseline model on various MOBO datasets. Each bar shows the RMSE resulting from zeroing out the corresponding frequency band in the AE input, with a band width of (a)~$10~\mathrm{kHz}$ and (b)~$2~\mathrm{kHz}$. }
\label{fig:ablation_freq_band}
}
\end{figure}

\subsubsection{AE scalogram frequency bands}
\label{subsubsec:abla_freq}
Next, inspired by Tempelman et al.~\cite{tempelman2022detection, tempelman2024uncovering}, we assessed the importance of individual frequency bands. As shown in Fig.~\ref{fig:ablation}, we partitioned the AE scalogram into overlapping frequency bands (\textit{e.g.},~$15\mathrm{-}20~\mathrm{kHz}$,~$25\mathrm{-}35~\mathrm{kHz}$,~etc.) with different widths. Each selected band was masked (replaced with zeros) to generate a series of mask-one-band-out (MOBO) datasets. These were used to evaluate our above pre-trained model with the SBD dataset, allowing us to quantify the impact of each band on validation performance without altering the CNN model. 

Following this approach, Fig.~\ref{fig:ablation_freq_band} presents the resulting relative MSE (RMSE) values, where the horizontal axis denotes the masked frequency range and the vertical axis indicates the corresponding RMSE. In this context, a higher RMSE suggests that the masked frequency band plays a more critical role in the prediction of KH porosity. Notably, Fig.~\ref{fig:ablation_freq_band}(a) reveals that the most important frequency bands, shown as $25\mathrm{-}45~\mathrm{kHz}$ in the red dashed box, align well with the critical frequency range previously identified by Tempelman et al.~\cite{tempelman2022detection}. 

To refine this further, we zoomed in on the frequency axis and masked narrower bands, creating a new series of the MOBO datasets. As shown in Fig.~\ref{fig:ablation_freq_band}(b), the model's validation results on the new MOBO datasets show that the $37\mathrm{-}45~\mathrm{kHz}$ band is the most critical under comparison. These frequencies align well with keyhole oscillation signatures reported in prior studies~\cite{khairallah2021onset}, suggesting that keyhole oscillation, as captured by AE, plays a critical role in KH porosity formation and its inference.

\section{A potential application}
\label{sec:app}
Building upon the results and analysis outlined above, we further investigated the potential application space enabled by our findings on AE-based KH porosity quantification. In this section, we propose and elaborate on a practical use case focused on AE-based porosity-informed process characterization and a potential idea for the reconstruction of the KH-process window boundary (KH-bound) for metallic LPBF AM. 

\begin{figure}[!ht]
\center{\includegraphics[width=\linewidth]
{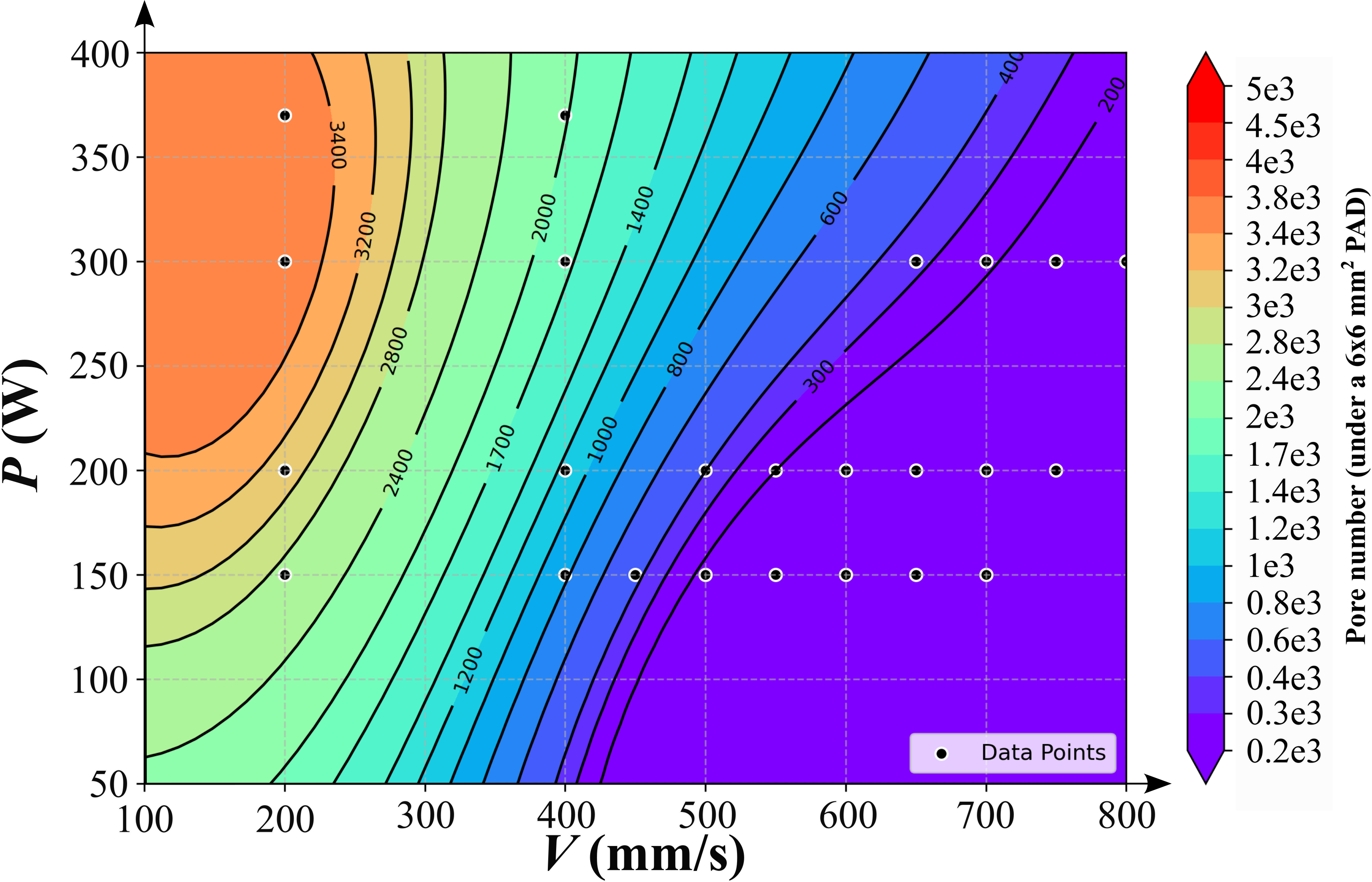}
\caption{Heat map and isocontours of $\texttt{KHNum}$ generated using the curated PAD dataset and the pre-trained baseline model. Note that the shape of the isocontours may vary depending on the distribution of selected $P\mathrm{-}V$ points. }
\label{fig:isocontour}
}
\end{figure}

By collecting and processing airborne ultrasonic AE at various grid points that span the KH and process window regimes on the $P\mathrm{-}V$ process map, we should be able to construct, using our proposed experimental and data-driven methodology, a KH porosity-related scalar field superimposed on the map for a given type of alloy. As one would reasonably anticipate, moving toward a zone with higher $P$ and lower $V$ (that is, toward the upper left region of the $P\mathrm{-}V$ space typically associated with elevated fusion energy density) will lead to intensified KH and higher levels of KH porosity. In contrast, KH porosity gradually decreases as the conditions shift toward regions of lower fusion energy density, indicating a smooth transition into stable conduction mode fusion. As such, this framework enables the generation of a continuous KH porosity heat map over the process parameter space, from which iso-value contours (isocontours) can be extracted to represent different severity levels of KH porosity generation. Following this idea, we constructed a level-set map for $\texttt{KHNum}$ (derivable from $\texttt{KHLineNum}$ given certain time and $V$) using our curated PAD dataset and pre-trained ML model, shown in Fig.~\ref{fig:isocontour}. 

We aim to illustrate that, by leveraging the AE scalogram in combination with the $V$ parameter to quantify KH porosity, the proposed approach offers a practical means of estimating the expected level of KH porosity associated with a selected $P\mathrm{-}V$ condition. This capability holds promise for guiding quality assurance and process parameter optimization in LPBF utilizing merely a few milliseconds of the airborne AE data. For instance, as shown in Fig.~\ref{fig:isocontour}, if a maximum allowable $\texttt{KHNum}$ of $300$ is imposed for a given PAD, the minimum acceptable $V$ can then be inferred to be approximately $400~\mathrm{mm/s}$, based on the intersection of the $\mathrm{KHNum}=300$ isocontour with the $V$-axis. 

\begin{figure}[!ht]
\center{\includegraphics[width=\linewidth]
{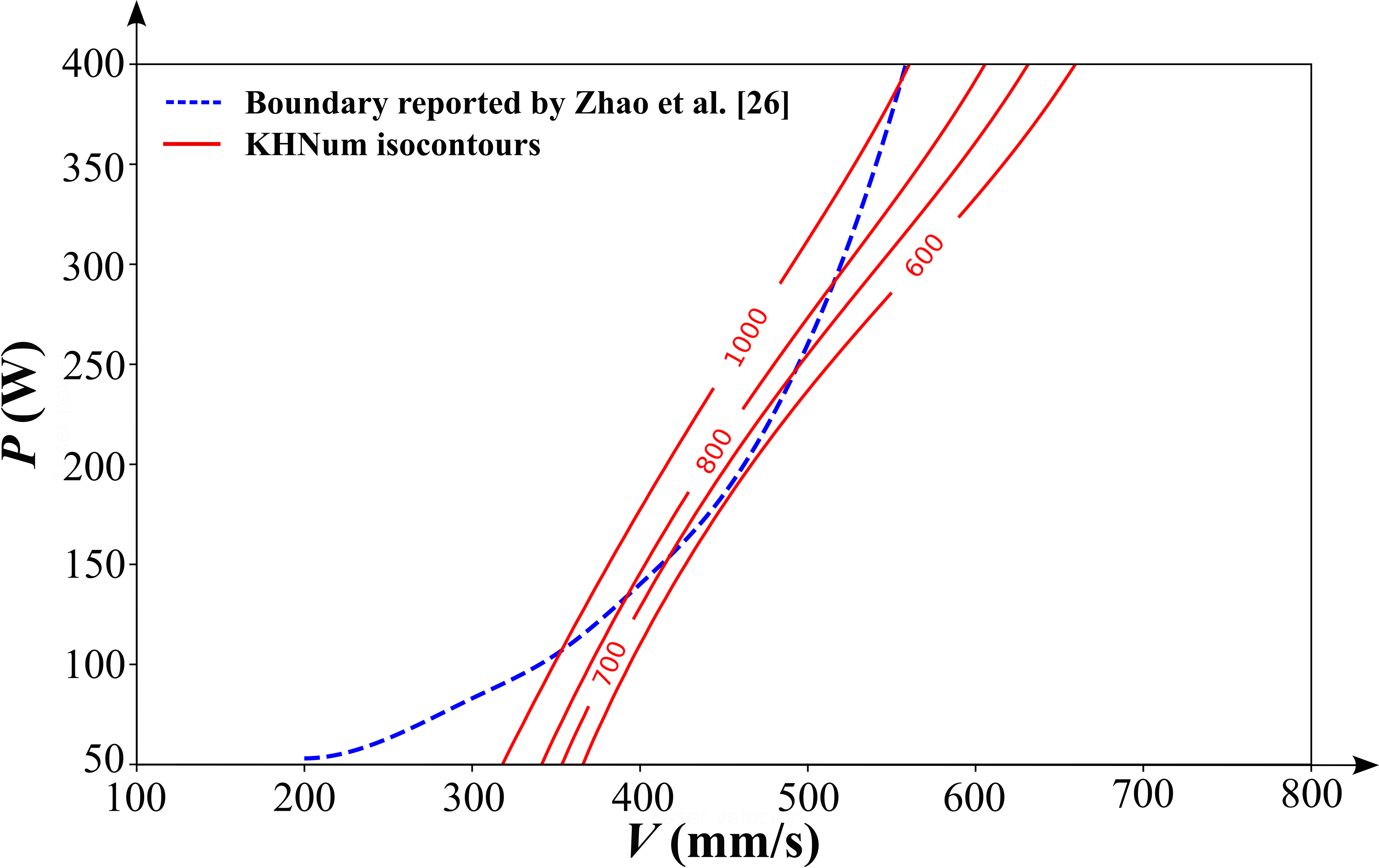}
\caption{AE-derived KH-bounds (the selected red $\texttt{KHNum}$ isocontours from Fig.~\ref{fig:isocontour}) and their comparison with the established KH-bound (the blue dashed line)~\cite{zhao2020critical}. The trend and location of reported KH-bound on the $P\mathrm{-}V$ process map match well with $\mathrm{KHNum}=600$~to~$\mathrm{KHNum}=1000$ isocontours for PAD prints. }
\label{fig:boundaries}
}
\end{figure}

Likewise, by selecting the $\mathrm{KHNum}=600$~to~$\mathrm{KHNum}=1000$ isocontours as representative boundaries between the KH and process window regimes for Ti-64 PAD, we observed in Fig.~\ref{fig:boundaries} a strong spatial alignment with the previously established KH-bound on the $P\mathrm{-}V$ process map, as reported in~\cite{zhao2020critical}. Although these $\texttt{KHNum}$ isocontours do not precisely capture the curvature of the ground-truth KH-bound---primarily due to the relatively coarse sampling resolution employed in our $P\mathrm{-}V$ grid design (Fig.~\ref{fig:PV_design})---the observed deviation remains within a practically acceptable margin. As such, we interpret these isocontours as \textit{AE-derived KH-bounds}: simplified yet effective empirical AE-informed approximations of the real experimental KH-bound. Their close alignment underscores the utility of our AE-based porosity quantification framework in delineating KH and process window regimes with minimal overhead and high spatial fidelity. While simplified, these AE-derived KH-bounds offer a promising surrogate to guide process parameter selection without requiring extensive XCT characterization. 

\begin{figure}[!ht]
\center{\includegraphics[width=\linewidth]
{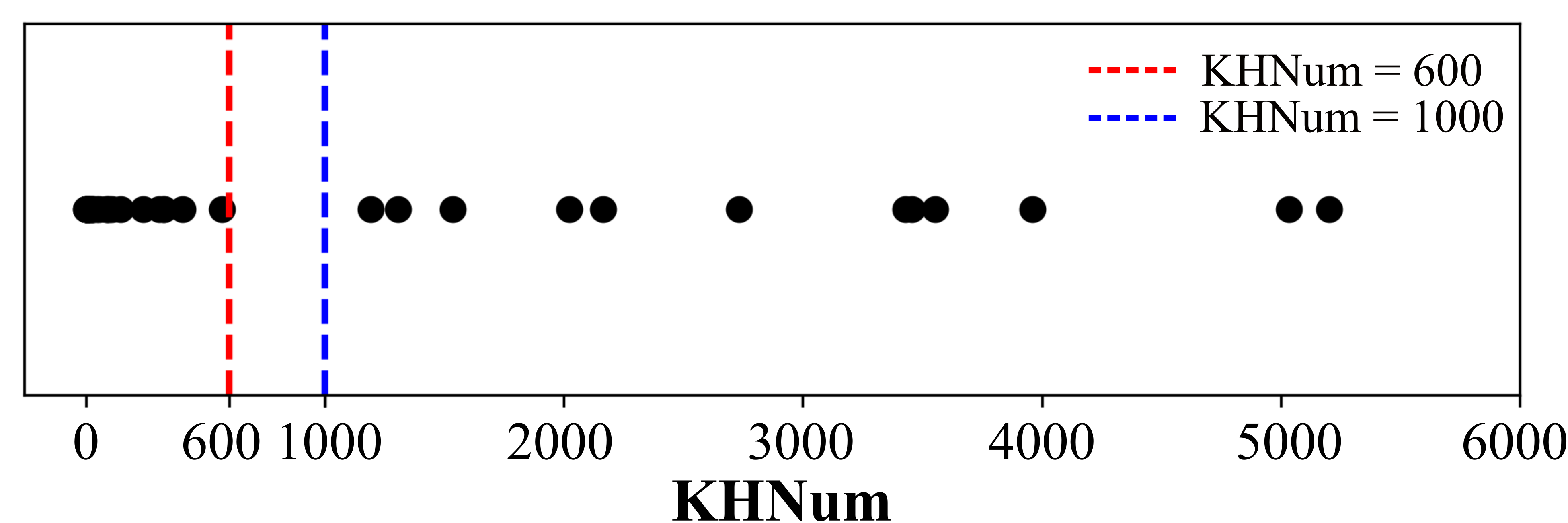}
\caption{1-D plot of $\texttt{KHNum}$ for all PADs. The gap between $600$ (red dashed line) and $1000$ (blue dashed line) matches our selection for the AE-derived KH-bounds. }
\label{fig:1D}
}
\end{figure}

We further examined the rationale behind selecting the $\mathrm{KHNum}=600$~to~$\mathrm{KHNum}=1000$ isocontours as the AE-derived KH-bounds. Prior work by Huang et al.~\cite{huang2022keyhole} reported that the frequency of keyhole fluctuations changes markedly as the $P\mathrm{-}V$ condition approaches the true KH-bound. This observation suggests that the heightened instability associated with such fluctuations facilitates KH pore formation, supporting the hypothesis that a sharp (as opposed to a gradual) rise in KH porosity may occur as the process transitions across the KH regime boundary. To test this hypothesis, we visualized $\texttt{KHNum}$ of the examined PAD builds in a 1-D plot. As shown in Fig.~\ref{fig:1D}, a pronounced increase in $\texttt{KHNum}$ is observed between $600$ (red dashed line) and $1000$ (blue dashed line), aligning closely with our chosen $\texttt{KHNum}$ isocontours. This consistency confirms that the AE-derived KH-bounds capture the underlying physics of keyhole oscillation and instability, which ultimately governs the onset of KH porosity as well as the true KH-bound.

If further validated across materials, this methodology may offer a foundation for generalizing AE-based KH porosity inference across alloy systems that exhibit similar keyhole dynamics. Given this, we propose the following application:

\textbf{Process map boundary reconstruction}---Assuming that our methodology captures the fundamental physics of liquid-vapor interplays, we can utilize our experimentally grounded framework for cost-effective reconstruction of the KH-bound in the $P\mathrm{-}V$ process map for an unknown alloy. The framework starts with a set of SBD experiments across selected $P\mathrm{-}V$ points, with sampling density adjusted based on the desired resolution of the process map. Airborne AE data are collected \textit{in~situ} during each print. Without requiring \textit{ex~situ} XCT-based porosity extraction or spatiotemporal registration, a pre-trained ML model (originally developed on a known alloy) is directly applied to the new AE data. Given an AE scalogram and its associated $V$ value, the model predicts a KH porosity measure (\textit{e.g.}, $\texttt{KHLineNum}$) at each $P\mathrm{-}V$ point. An empirical or numerical threshold on this measure for the designated SBDs can then be used to delineate an approximate KH-bound. This approach leverages AE and generalizable ML inference to avoid the cost and complexity of XCT-based characterization.

\section{Discussion}
\label{sec:discussion}
Beyond the technical aspects of this study, we provide a more in-depth discussion regarding the interpretation of our results, linkages between our work and others' findings, limitations associated with the presented approach, and potential future work for the next steps. 

First, we explore the underlying physical mechanisms that support AE-based inference of $\texttt{KHLineNum}$ using a relatively straightforward ML framework. As evidenced by previous studies~\cite{ren2024sub}, it is widely recognized that airborne AE signals are strongly influenced by the dynamical morphologies of the keyhole, melt pool, and plume. We find that the spectral characteristics of AE are primarily governed by the transient oscillation modes of the keyhole, which are intrinsically tied to the transient dynamics of keyhole formation, collapse, and the eventual development of KH pores, as well as $V$ which determines the rate of material melting and vaporization. This interpretation aligns well with findings reported by Tempelman et al.~\cite{tempelman2024uncovering}, as well as our own results, which consistently highlight prominent AE frequency bands that match known keyhole oscillation frequencies. These observations suggest that keyhole instability coupled with certain laser scanning speeds may emit distinctive AE signatures, serving as precursors to KH porosity formation events and enabling AE's capability of $\texttt{KHLineNum}$ quantification along the scanlines. 

Second, as illustrated in Fig.~\ref{fig:flowchart}, we conceptualize the essence of all physical observations and measurable data in LPBF---including AE, porosity, keyhole and melt pool dynamical behaviors, etc.---as different manifestations of common driving sources: the spatiotemporally varying fusion energy densities and transient laser-melt interplays. These shared origins imply a set of latent relationships among these phenomena; in the specific case where KH pores are generated, such energy inputs and interplays tend to be excessive, necessitating more sophisticated methods for precise characterization. While analytically or numerically capturing such multiphysics interactions remains a formidable challenge, our results suggest that a rather tractable data-driven approach, together with airborne AE and an appropriate porosity measure ($\texttt{KHLineNum}$), provides a practical and effective pathway to uncover and quantify these connections. Based on the findings and insights provided by Tempelman et al.~\cite{tempelman2022detection, tempelman2024uncovering} and Ren et al.~\cite{ren2024sub}, we advance the methodology by developing a regression framework that quantifies KH porosity (in the form of $\texttt{KHLineNum}$) directly from airborne AE spectral-temporal representations and corresponding process parameters. As such, we believe this contribution not only strengthens the feasibility of cost-efficient AE-based process monitoring highlighted by previous work but also opens new avenues for exploring the fundamental physics of laser-material interaction and the mechanisms linking keyhole instability, pore formation, and AE signal generation. 

Third, we acknowledge several limitations inherent to this study. First and foremost, the applicability of the proposed AE-based sensing approach may diminish significantly in LPBF systems employing multiple lasers, where signal superposition and source ambiguity can obscure the origin of detected acoustic events. Furthermore, all experiments in this work were carried out within a relatively confined spatial domain (maximum dimension~$<10~\mathrm{mm}$), which may limit the generalizability of the trained model to larger-scale builds that span the entire build plate. To begin addressing these constraints, future work may explore the deployment of multiple AE sensors strategically positioned at different places and directions to improve localization and source separation. Aside from the above scaling issues, we also realized that the potential for material generalizability, hypothesized and discussed in Sec.~\ref{sec:app}, still remains unverified. Demonstrating the applicability of the proposed AE-based KH porosity quantification framework across different materials is essential for fully realizing its broader utility in process map reconstruction. Moreover, another non-negligible limitation arises from the intrinsic multilayer nature of 3D printing. In practical LPBF processes, remelting of underlying layers during subsequent scans can lead to partial or complete elimination of previously formed pores. While our framework is capable of detecting and quantifying KH pore formation \textit{in~situ}, it does not account for post-formation pore dynamics such as remelting-induced elimination. Prior work by Van Petegem et al.~\cite{de2024healing} has suggested the feasibility of detecting individual pore elimination events using AE sensing as well, which may offer a promising opportunity for future integration into part-scale porosity tracking. Incorporating such capabilities would enhance the temporal completeness of AE-based porosity inference, bridging both formation and removal events. 

Lastly, we anticipate a broader application space stemming from the proposed methodology. For instance, prior work by Reddy et al.~\cite{reddy2024fatigue} has demonstrated a strong correlation between the total number density of porosities and the fatigue life of as-built components. This finding underscores the potential to exploit the quantification of KH pore count based on AE as a predictive tool for the evaluation of fatigue performance. Furthermore, with AE sensors possessing sufficient bandwidth and spatial sensitivity, it is conceivable to detect the KH porosity generation rate over time windows as short as a few milliseconds. Such high-resolution feedback may serve as a basis for future efforts in rapid process–property correlation, and with further development, could contribute to closed-loop control strategies.

\section{Conclusion}
\label{sec:conclusion} 
This study presents a framework for quantitatively inferring keyhole porosity in metallic LPBF using airborne AE sensing. Leveraging a series of carefully controlled bare-plate experiments, we achieved sub-scanline data synchronization by spatiotemporally registering \textit{in~situ} AE scalogram snippets, spectrotemporal representations of airborne AE data generated via CWT, with corresponding \textit{ex~situ} XCT porosity measurements at the millisecond scale. Central to our approach is a spatially resolved porosity metric named $\texttt{KHLineNum}$, defined as the number of KH pores per unit scan length, which outperforms other traditional KH porosity measures in AE-based porosity regression tasks. Using $\texttt{KHLineNum}$, we trained a tractable CNN that accurately quantifies KH porosity in a short duration, generalizes across varying scan geometries and parameters, and enables short-time porosity prediction. Moreover, our framework facilitates the inference of KH regime boundaries in the $P\mathrm{-}V$ space directly from AE data, demonstrating its potential for rapid process characterization and control. These findings advance the development of AE-driven, quantitative process monitoring in LPBF, with promising implications for adaptive process control and optimization in AM.

\section*{Acknowledgement}
\label{sec:acknow}
This research is supported by the Eaton Corporation Award Number 001145471. We thank Dr.~Alexander Myers and Daniel Diaz for useful discussions. We thank Dr.~Christian Gobert for the development of data acquisition software. We thank Doug Washabaugh, Liza Allison, Dr.~Sandra DeVincent Wolf, Dr.~Nicholas Lamprinakos, Milly Yi, and Zhaoxuan Ge for their support and guidance on the LPBF experiments. Xuzhe Zeng thanks Buqing Ou for her comments on the draft and help with figure layout. Haolin Liu thanks Silin Liu for the help with paper proofreading and figure generation. 

\bibliographystyle{elsarticle-num-names}
\bibliography{references}

\clearpage

\appendix

\section{Additional results from qualitative correlation analysis between the AE scalograms and the KH porosity metrics}
\label{appendix:qual_addi}
As mentioned in Sec.~\ref{subsec:qualitative_analysis}, we hypothesized that adding $P$ as a separate channel to the current input tensor (size: $224\times 224\times 3$) would enhance the model's predictive fidelity. An ML model with the same architecture and training scheme was trained using the same portion of the SBD dataset. We then validated these models on the SBD validation dataset, and generated the $R^2$ plots in a way similar as Sec.~\ref{subsec:pred_acc}. 

\begin{figure}[!ht]
\center{\includegraphics[width=0.9\linewidth]
{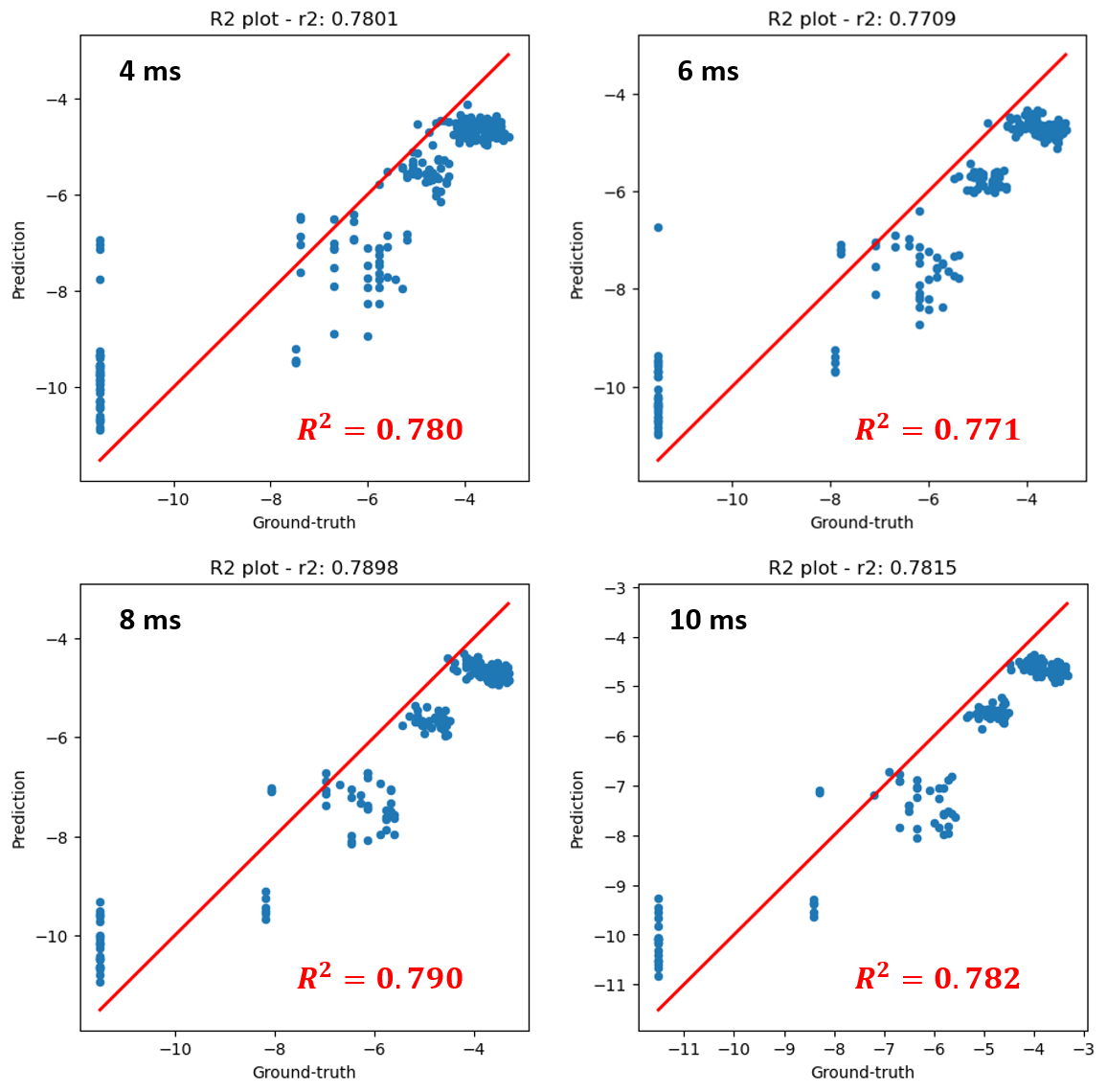}
\caption{$R^2$ plots of SBDs with $P$ as an additional input channel. }
\label{fig:r2_SBD_withP}
}
\end{figure}

As shown in Fig.~\ref{fig:r2_SBD_withP}, contrary to our expectations, the inclusion of $P$ as an additional input channel led to a noticeable decline in model performance across all snippet duration groups, compared with the results reported in Sec.~\ref{subsec:pred_acc}. The underlying reasons for this unexpected performance degradation remain unclear to the authors and need further investigation. 

\section{Comparative study of ML architectures based on $R^2$ scores}
\label{appendix:ML_models}
Although our proposed framework employs a lightweight CNN model, we also evaluated several alternative ML architectures and compiled their $R^2$ scores for comparison. Table~\ref{tab:ML_models} summarizes the detailed $R^2$ results for PAD validation. Both the ResNet \cite{he2016deep} and ViT \cite{dosovitskiy2020image} models were fine-tuned from pretrained weights---specifically, ResNet-18 and ResNet-34 for the ResNet variants, and the ViT-Base/16 model (officially referred to as \texttt{vit-base-patch16-224}) for the ViT. The validation results on the PAD dataset indicate that all models achieve strong predictive performance, with consistently high values of $R^2$. This outcome underscores the robust correlation between acoustic signals and porosity, a central finding of this study. In particular, we demonstrate that even a relatively simple architecture achieves high $R^2$ values in validation, underscoring the intrinsic relationship between acoustic emissions and the resulting porosity. Increasing the complexity of the model yields only marginal improvements in the precision of KH porosity quantification. Consequently, the CNN2D architecture is selected, as it offers an optimal balance between predictive accuracy and computational efficiency, thus streamlining the overall pipeline. 
\begin{table}[!ht]
    \centering{
    \caption{$R^2$ performance comparison across different models for the validation of PADs}
    \vspace{-2mm}
    \resizebox{\linewidth}{!}{
    \begin{tabular}{c|c|c|c|c}
        \toprule
        \backslashbox{\textbf{Snippet~(ms)}}{\textbf{Model}} & \textbf{CNN2D} & \textbf{ResNet18} & \textbf{ResNet34} & \textbf{ViT} \\
        \midrule
        \textbf{$4$} & $0.913$ & $0.862$ & $0.894$ & $0.943$ \\ 
        \midrule
        \textbf{$6$} & $0.903$ & $0.854$ & $0.903$ & $0.910$ \\ 
        \midrule
        \textbf{$8$} & $0.941$ & $0.848$ & $0.912$ & $0.938$ \\
        \midrule
        \textbf{$10$} & $0.939$ & $0.838$ & $0.912$ & $0.945$ \\
        \bottomrule
    \end{tabular}}
    \label{tab:ML_models}}
\end{table}

\end{document}